\documentclass[a4paper,11pt]{article}
\pdfoutput=1 

\usepackage{jcappub} 
\usepackage{amsmath}
\usepackage{slashed}
\newcommand{\be}{\begin{equation}}
\newcommand{\ee}{\end{equation}}
\newcommand{\bea}{\begin{eqnarray}}
\newcommand{\eea}{\end{eqnarray}}

\title{Reviewing the prospect of fermion triplets as dark matter and source of baryon asymmetry in non-standard cosmology}

\author[a, b]{Anirban Biswas,}
\author[c]{Mainak Chakraborty,}
\author[d]{Sarif Khan}

\affiliation[a]{Department of Physics \& Lab of Dark Universe, Yonsei University,
50 Yonsei-ro, Seodaemun-gu, Seoul 03722, South Korea}
\affiliation[b]{Center for Quantum Spacetime, Sogang University,
35 Baekbeom-ro, Mapo-gu, Seoul 121-742, South Korea}
\affiliation[c]{ School of Physical Sciences, Indian Association for the Cultivation of Science, 2A $\&$ 2B Raja S.C. Mullick Road, Kolkata 700 032, India }
\affiliation[d]{Institut f\"{u}r Theoretische Physik, 
Georg-August-Universit\"{a}t G\"{o}ttingen, Friedrich-Hund-Platz 1,
37077 G\"{o}ttingen, Germany}

\emailAdd{anirban.biswas.sinp@gmail.com}
\emailAdd{psmc2382@iacs.res.in}
\emailAdd{sarif.khan@uni-goettingen.de}

\abstract{
Indirect searches of Dark Matter (DM), in conjugation with `missing track searches' at the collider seem to confine SU(2)$_L$ fermion triplet DM (FTDM) mass within a narrow range around 1 TeV. The canonical picture of the pure FTDM is in tension since it is under-abundant for the said mass range. Several preceding studies have reported that an extra species ($\phi$), redshifts faster than the radiation ($\sim a^{-(4+n)}$ where $n>0$), leads to a faster expanding early Universe by dominating in the energy density with an enhanced Hubble parameter. This has the potential to revive the under-abundant FTDM ($\mathbb{Z}_2$ odd, lightest generation) by causing freeze-out earlier without modifying the interaction strength between DM and thermal bath. On the other hand, although the CP asymmetry produced due to the decay of $\mathbb{Z}_2$ even heavier generations of the triplet remains unaffected, its evolution is greatly affected by the non-standard cosmology. It has been observed through numerical estimations that the minimum mass of the triplet, required to produce sufficient baryon asymmetry of the Universe (BAU), can be lowered up to two orders (compared to the standard cosmology) in this fast expansion scenario. The non-standard parameters $n$ and $T_r$ (a reference temperature below which radiation dominance prevails), which simultaneously control DM abundance as well as the frozen value of BAU, are tightly constrained from the observed experimental values. We have found that $n$ is strictly bounded within the interval $0.4\lesssim n \lesssim 1.8$ where the upper bound is imposed by the BAU constraint whereas the lower bound arises to satisfy the correct DM abundance. It has been noticed that the restriction on $T_r$ is not so stringent as it can vary from sub GeV to a few tens of GeV.
}

\makeatletter
\gdef\@fpheader{}
\makeatother

\begin{document}
\maketitle
\flushbottom
\newpage
\section{Introduction} 
\label{Intro}
The origin of dark matter(DM) and the explanation of the observed
matter-antimatter asymmetry are two most important and unresolved
issues in particle and astroparticle physics. The existence
of dark matter has been suggested by a plethora of observations at the
galactic scale like galactic rotation curves \cite{Sofue:2000jx}, gravitational lensing
of distant celestial bodies \cite{Bartelmann:1999yn}, colliding galaxy clusters
(e.g. the Bullet cluster) \cite{Clowe:2003tk} etc.
Similarly, the observations at the cosmological scale by
WMAP \cite{WMAP:2012fli} and Planck \cite{Planck:2015fie,Planck:2018vyg} satellites have measured most precisely the abundance of dark matter till date,
$\Omega_{\rm DM} h^{2}=0.120\pm0.001$ at 68\% C.L. \cite{Planck:2018vyg}, 
which is roughly around 25\% energy budget of our Universe.
While Weakly Interacting Massive Particle (WIMP)
\cite{Silveira:1985rk, Scherrer:1985zt, Srednicki:1988ce, Gondolo:1990dk} characterised by 
the GeV--TeV scale mass window with weak scale interaction cross-section
is one of the most preferred dark matter candidates, there exist
various other possibilities like, non-thermal dark matter
\cite{Hall:2009bx, Biswas:2015sva, Bernal:2017kxu},
asymmetric dark matter \cite{Petraki:2013wwa},
primordial black hole as dark matter \cite{Carr:2021bzv} etc.  
On the other hand, there are overwhelming evidences
in support of excess baryonic matter over anti-matter at
various cosmological epochs. The baryon-to-photon ratio
($\eta_B$) predicted during the Big bang nucleosynthesis (BBN) by measuring
various light element abundances matches excellently with that obtained
from the baryon density at the time of recombination. The current acceptable
value of $\eta_B$ which carries the information about the baryon asymmetry\footnote{It is to be noted that baryon asymmetry can also be expressed as net baryon number density scaled by entropy density
i.e $Y_B=\frac{n_B}{s}$ and its experimental value lies within $(9.05 \pm 0.150 )\times 10^{-11}$. }
of the Universe (BAU) is $\eta_B= (6.4 \pm 0.1 )\times 10^{-10}$
at 95\% C.L. \cite{Planck:2018vyg}. Non-zero neutrino masses and large leptonic mixing \cite{Pontecorvo:1967fh,Gribov:1968kq} have now become experimentally \cite{Abe:2017vif,Abe:2018wpn,Adamson:2017gxd,NOvA:2018gge,Adey:2018zwh,Bak:2018ydk,Adamson:2013whj,Abe:2014bwa} 
verified fact and the experiments are now inching towards
precise measurement of the oscillation parameters \cite{Esteban:2018azc}. Various seesaw mechanisms \cite{Minkowski:1977sc,GellMann:1980vs,Yanagida:1980xy,Mohapatra:1979ia,Foot:1988aq,Ma:1998dn}, introduced to explain tiny mass of the active neutrinos, can act as a bridge between
DM phenomenology and mechanism of dynamical generation of baryon asymmetry through 
lepton asymmetry(popularly known as Baryogenesis via Leptogenesis \cite{Fukugita:1986hr}).

In this work, we have considered one of the minimal extensions of
the SM to address both the dark matter and the baryon asymmetry of
the Universe. Here we have added three SU$(2)_{\rm L}$ fermionic triplets
with hypercharge $Y=0$ in the particle spectrum of the SM. As a result,
two of the three light neutrinos have sub-eV Majorana masses through the
Type-III seesaw mechanism and the remaining one continues to be massless
due to the imposition of an additional $\mathbb{Z}_2$ odd parity on the
lightest triplet fermion $\Sigma_{R\,1}$. Accordingly, the triplet
$\Sigma_{R\,1}$ gets decoupled from the neutrino sector and its neutral component
$\Sigma^0_1( = \Sigma^0_{R\,1} + {\Sigma^0_{R\,1}}^c)$ could be a
Majorana type dark matter candidate which satisfies the relic density
bound for $M_{\Sigma_1} \simeq 2.3$ TeV \cite{Ma:2008cu}.
However, such a dark matter candidate in the mass range $1.5\,\,{\rm TeV}\,\,
\lesssim M_{\Sigma_1} \lesssim  2.7$ TeV
is already ruled out by the indirect detection bound coming from dark matter
annihilation into $W^{+} W^{-}$ and $\gamma\gamma$ final states
\cite{Hisano:2003ec,Hisano:2004ds}. There are some earlier attempts to generate
correct relic abundance for the low mass triplet fermionic dark matter 
by either introducing an additional non-thermal contribution to the relic density
from a decaying scalar \cite{Biswas:2018ybc} or including a singlet fermion
with a nonzero mixing with the triplet such that the resulting dark matter
candidate is predominantly a singlet-like state \cite{Choubey:2017yyn}. 
In this work, we are considering another interesting way to bring
relic density in the correct ballpark value of Planck collaboration \cite{Planck:2018vyg}
in the low mass regime of the triplet dark matter.
Here, we modify the underlying
cosmology of the early\footnote{The word `early' signifies an era which started earlier than the thermal production scale of heavy fermion triplets and lasts till the freeze-out epoch of the DM.} Universe by
adding extra species which redshifts faster than the
radiation content of the Universe.
There is a large ambiguity about the cosmology of the University
prior to the era of light nuclei (H, D, $^{3}$He, $^4$He, and $^7$Li)
formation, popularly known as
the Big bang nucleosynthesis which began when the age
of the Universe was $\sim 1$ second and it continued for about
three minutes. More importantly, the theoretical predictions
of BBN (e.g. light nuclei abundance), considering radiation dominated
Universe, are in good agreement with observational data. On the contrary,
very little information is available to us about the era before BBN and in most of
the literature a radiation dominated Universe has been considered. 
However, the expansion rate of the Universe driven by the Hubble
parameter can be higher or lower compared to the case of radiation
domination depending on how the total energy density of the Universe
redshift with respect to expansion. The fast expansion results if
there is an additional source of energy on top of the radiation
background composed of the SM particles such that the total
energy density $\rho_{\rm tot}>\rho_{\rm rad}$, the energy density
due to radiation. The early matter domination \cite{Cosme:2020mck,
Evans:2019jcs, Biswas:2022fga, Borah:2022byb}, the kination \cite{Tsujikawa:2013fta, Ferreira:1997hj, Gouttenoire:2021jhk, Salati:2002md,Profumo:2003hq,DEramo:2017gpl}
etc. are a few examples of fast expanding Universe. On the other hand,
the opposite situation could also arise particularly
in some modified gravity theory \cite{Nojiri:2010wj, Catena:2007ix, Okada:2009xe, Meehan:2014bya, Nojiri:2017ncd} where a negative contribution in the energy density leads to
slow expansion \cite{Biswas:2023eju}. 

In this work, we consider a fast expanding Universe
by introducing a species $\phi$ that dominates the energy
density in the early Universe. Moreover, the energy density of
$\phi$ red-shifts at a faster rate with the cosmic scale factor
($a$) compared to that of the radiation ($\rho_{\rm rad}\propto a^{-4}$) i.e. $\rho_{\phi}\propto a^{-(4+n)}$ with $n>0$. Therefore, although initially we
have $\phi$ dominance, the radiation eventually takes over from $\phi$ before
the start of BBN ($T\sim 1$ MeV) since the latter decreases rapidly due to expansion. 
The revival of radiation domination is characterised by the temperature $T_r\geq 1$ MeV.
Accordingly, we have two new parameters $n$ and $T_r$ (both are not independent)
controlling the cosmology of the Universe for $T>T_r$. Depending upon the value of
$n$ we can have different expansion histories of the Universe. For example, the scenario
with $n=2$ can be realised if the kinetic energy of a scalar field dominates the
energy budget, which is known as the Kination era where
$\rho_{\phi} \propto a^{-6}$ \cite{Gouttenoire:2021jhk, Salati:2002md, Profumo:2003hq, DEramo:2017gpl}. This simple model for non-standard
cosmology was first proposed in \cite{DEramo:2017gpl} and thereafter a number of works have been done using this model in the context of dark matter
\cite{DEramo:2017ecx, Biswas:2018iny, Ganguly:2022qxs, Ghosh:2022fws, Arcadi:2021doo, Barman:2021ifu} and
leptogenesis \cite{Chen:2019etb, Mahanta:2019sfo,Konar:2020vuu,Chang:2021ose,Chakraborty:2022gob, Mahanta:2022gsi, DiMarco:2022doy}.
In this work, our prime motivation is to find an allowed parameter space for the pure
triplet dark matter which otherwise is entirely ruled out by the combination of relic density and indirect detection bounds for the standard cosmological scenario. In the
radiation dominated Universe, a triplet fermion of mass lower than 2.3 TeV remains
under abundant due to large annihilation and co-annihilation cross-sections into   
electroweak gauge bosons and fermions, which results in a late freeze-out. However,
this situation can be changed in presence of a fast expanding Universe characterised
by a Hubble parameter larger than the standard scenario. Moreover, the non-standard
cosmology also affects the final baryon asymmetry produced from lepton
asymmetry through sphaleron involving processes before the electroweak phase transition.
In this model the other two heavier triplet fermions, which are even under $\mathbb{Z}_2$
symmetry, have non-vanishing Yukawa couplings with Lepton doublets and Higgs doublet.  
Accordingly, the lepton asymmetry can be generated due to CP violating out of equilibrium decays of these heavy triplet fermions at very high temperature. This has been studied earlier considering a radiation dominated Universe before the BBN and found \cite{Hambye:2003rt,AristizabalSierra:2010mv,Hambye:2012fh} that the
minimum mass of triplet fermion required to generate the observed baryon asymmetry
is $M_{\Sigma} \gtrsim 3\times 10^{10}$ GeV. In the present work, we explore the
possibility of relaxing the above mentioned lower
bound\footnote{One can also lower the triplet mass bound by extending the minimal
model with additional fields as done in \cite{Vatsyayan:2022rth}.} on $M_{\Sigma}$
obtained from BAU bound. Predominance of the
extra scalar field $\phi$ in the primordial Universe resulted in faster (than standard cosmology)
expansion of the Universe. Thus enhancement caused in the Hubble parameter pushes the rate of decay/Inverse decay (of $\Sigma_2$ or $\Sigma_3$) and scattering towards a lower value. Therefore 
the Boltzmann equations governing the evolution of produced asymmetry will be altered with respect to its standard counterpart. Although the modification of cosmology doesn't affect the asymmetry production, it has non-trivial effect in the evolution of the same. Our aim in this part of the work is to identify the specific region of the parameter space which can produce observed baryon asymmetry below the limit (on the decaying triplet mass) set using standard cosmology.

The fascinating feature of this study is that the newly introduced non-standard cosmology 
co-relates the apparently decoupled dark sector and the Yukawa sector(responsible CP asymmetry generation) through the non-standard parameters $T_r,n$. We expect that the relic density bound together with the observed BAU bound will allow a strongly constrained parameter space spanned by the yet unbounded non-standard parameters $T_r - n$.

To address the issues raised in the preceding paragraphs in a comprehensive and systematic manner, the rest of this paper is organized as follows. The minimal extension scheme of SM adopted here, to account for the DM relic abundance as well as the BAU, is discussed in Sec.\,\,\ref{model}. A qualitative overview about the generation of active neutrino mass and mixing through Type-III seesaw mechanism is presented in Sec.\,\,\ref{seesaw}. Sec.\,\,\ref{trip-dm} is solely devoted to the study of DM-phenomenology. Its different subsections are arranged chronologically so that the need for introduction of non-standard cosmology to save the fermion triplet DM scenario can be clearly understood. So we start with a comprehensive analysis of existing bounds (obtained from indirect searches and collider experiments) on fermion triplet DM in sec.\,\,\ref{cons-DM}. The collective results allow only a window for triplet DM mass, which is not enough to produce adequate relic abundance in standard cosmological scenario. Thus non-standard cosmology is introduced in Sec.\,\,\ref{NS_cosmo} and its subsections discuss how the freeze-out temperature of the DM can be modified (Sec.\,\,\ref{mod-tf}) in the newly introduced cosmology, without harming the successful predictions of the BBN (Sec.\,\,\ref{BBN-bound}). Sec.\,\,\ref{DM-num} presents a detailed account of numerical analysis related to the study of DM phenomenology in modified cosmological scenario. Sec.\,\,\ref{baryo-lepto} contains a rigorous analysis of Baryogenesis through Leptogenesis in the present scenario. Discussions about Leptogensis, continue in its constituent subsections, starting with generation of CP asymmetry due to the decay of next to lightest triplet (Sec.\,\,\ref{cpasy}), followed by a thorough description of the Boltzmann equations in the non-standard (Sec.\,\,\ref{boltzeqn}) cosmology and ends with a meticulous analysis (Sec.\,\,\ref{numboltz}) of numerical results supplemented by plots of suitable model parameters. The noteworthy outcomes of this exhaustive study are summarised in Sec.\,\,\ref{summary}. 
\section{Model}
\label{model}
In the present work, we have addressed a few well-recognised SM drawbacks namely dark matter, 
neutrino mass and matter anti-matter asymmetry of the Universe. To this end, we have extended
the SM particle content by three additional $SU(2)_{L}$ triplet fermions similar to the 
Type-III seesaw mechanism. The complete particle spectrum with the associated charge assignment 
is shown in Table \ref{tab1}.
Among them, the lightest one is $\mathbb{Z}_{2}$ odd and its neutral component serves 
as the suitable DM candidate whereas the other two are $\mathbb{Z}_{2}$
even and contribute to the neutrino mass through the Type-III seesaw mechanism. 
We have considered its $(\Sigma_1)$ mass in 
the TeV range. Fermion triplet DM heavier than $2.4$ TeV is ruled out by Planck data due to the over-abundance. Moreover, for mass beyond $1.5$ TeV, DM annihilation faces Sommerfeld enhancement \cite{Hisano:2003ec,Hisano:2004ds} which is in 
direct contradiction with the indirect detection bound provided by Fermi-LAT 
data \cite{Fermi-LAT:2009ihh}. 
Therefore, we have considered $\Sigma_{1}$ mass always below $1.5$ TeV. 
This mass restriction forbids the lightest fermion triplet to account for the full relic abundance of DM as reported by the experiments. We also discuss in the result section, how the presence of an extra species (denoted as $\phi$)
in the early Universe assist us to obtain the total amount of dark matter relic density
even below $1$ TeV, which was not possible in standard cosmological scenario. The other two triplet fermions, which are even under $\mathbb{Z}_{2}$ symmetry,
take part in the generation of neutrino mass, 
lepton asymmetry, are kept at a very high scale $\mathcal{O}(10^{9})$ GeV. 
The improvement of the lower bound on triplet mass, to obtain the correct value of BAU, 
due to the presence of the extra species in the early Universe is emphasized in the section dedicated to `Baryogenesis through Leptogenesis'.
\begin{center}
\begin{table}[h!]
\begin{tabular}{||c|c|c|c||}
\hline
\hline
\begin{tabular}{c}
    Gauge\\
    Group\\ 
    \hline
    
    ${\rm SU(2)}_{\rm L}$\\  
    \hline
    ${\rm U(1)}_{\rm Y}$\\ 
    \hline
    $\mathbb{Z}_{2}$\\ 
\end{tabular}
&

\begin{tabular}{c|c|c}
    \multicolumn{3}{c}{Quarks}\\ 
    \hline
    $Q_{L}^{i}=(u_{L}^{i},d_{L}^{i})^{T}$&$u_{R}^{i}$&$d_{R}^{i}$\\ 
    \hline
    $2$&$1$&$1$\\ 
    \hline
    $1/6$&$2/3$&$-1/3$\\ 
    \hline
    $1$&$1$&$1$\\ 
\end{tabular}
&
\begin{tabular}{c|c|c|c}
    \multicolumn{4}{c}{Leptons}\\
    \hline
    $L_{L}^{i}=(\nu_{L}^{i},e_{L}^{i})^{T}$ & $e_{R}^{i}$ 
    & $\Sigma^{c}_{Rj}$
    & $\Sigma^{c}_{R1}$\\
    \hline
    $2$&$1$&$3$&$3$\\
    \hline
    $-1/2$&$-1$&$0$&$0$\\
    \hline
    $1$&$1$&$1$&$-1$\\
\end{tabular}
&
\begin{tabular}{c}
    \multicolumn{1}{c}{Scalar}\\
    \hline
    $\Phi$\\
    \hline
    $2$\\
    \hline
    $1/2$\\
    \hline
    $1$\\
\end{tabular}\\
\hline
\hline
\end{tabular}
\caption{Particle contents and their corresponding
charges under the SM gauge group. }
\label{tab1}
\end{table}
\end{center}
The Lagrangian for the particle spectrum shown
in Table \ref{tab1} is given by,
\begin{eqnarray}
\mathcal{L}&=&\mathcal{L}_{\rm SM} + \mathcal{L}_{\Sigma}\,, 
\label{lag}
\end{eqnarray}
where $\mathcal{L}_{\rm SM}$ is the Lagrangian associated with the SM
particles and $\mathcal{L}_{\Sigma}$ is the Lagrangian for the triplet fermions which has the following form,
\begin{eqnarray}
\mathcal{L}_{\Sigma}&=&\frac{i}{2} \sum_{j=1,2,3}  
{\rm Tr} \left[ \overline{\Sigma_{R\,j}}\,\slashed{D} \Sigma_{R\,j}
+ \overline{{\Sigma_{R\,j}}^c}\,\slashed{D}{\Sigma_{R\,j}}^c \right]
-  \sum^{3}_{\alpha = 1} \sum^{3}_{j = 1} \left(\sqrt{2}\,
y^{\alpha j}_{\Sigma} \overline{L_{L\,\alpha}} \,\Sigma_{R\, j} \tilde{\Phi} 
+ h.c. \right) \nonumber\\
&-&\frac{1}{2}\sum^{3}_{j=1}  {\rm Tr} \left[\overline{{\Sigma_{R\,j}}^c} \left(M_\Sigma \right)_{jj} \Sigma_{R\,j}  +h.c \right],
\label{lagN}
\end{eqnarray}
where the above three terms are, respectively, the kinetic term, the Yukawa coupling term and the bare mass term for the fermion triplets $\Sigma_{j}$ ($j=1,2,3$). The Latin index $j=1\,$to 3 stands for three generations of fermion triplets whereas the Greek index $\alpha$ is used to designate three different lepton flavours. It is to be noted that although the summation on $j$ runs from $1$ to $3$ in the second term, the species $\Sigma_{R\,1}$ does not couple to the SM Higgs and the lepton doublets, which implies that the first column of the Yukawa coupling matrix $y_\Sigma$ is null. This has been
implemented by imposing a $\mathbb{Z}_2$ odd parity
to the lightest triplet generation {\it i.e.} $\Sigma_{R1}$ whereas all the other species are $\mathbb{Z}_2$ even including the SM particles.
It should be noted that we have constructed the physical basic using a linear combination of component field and its CP conjugate of each triplet $\Sigma_{Rj}$.
In the fundamental representation of SU(2)$_{\rm L}$, the triplet $\Sigma_{Rj}$
can be expressed as
\begin{equation}
\Sigma_{Rj }=\left(\begin{array}{cc}
         {\Sigma^0_{Rj}}/{\sqrt{2}} & \Sigma^+_{Rj} \\
         \Sigma^-_{Rj}   & -{\Sigma^0_{Rj}}/{\sqrt{2}}
        \end{array}\right)~.
 \end{equation}
 The $2\times 2$ representation of $\Sigma_{Rj}$ transforms under SU(2)$_{\rm L}$
 as $\Sigma^{\prime}_{Rj} = U \Sigma_{Rj} U^\dagger$ with $U = \exp(-i\frac{g}{2} {\theta_a\,\sigma_a})$ and $\sigma_a\,(a=1,\,2,\,3)$ are the Pauli matrices. The
 complex conjugate of $\Sigma_{Rj}$, ${\Sigma_{Rj}}^*$, does not transform in
 the similar manner as $\Sigma_{Rj}$ under SU(2)$_{\rm L}$. However, the combination $i\sigma_2\,{\Sigma_{Rj}}^{*}\,i\sigma_2$ has the identical transformation property as that of the triplet $\Sigma_{Rj}$. Therefore, in order
 to have an SU(2) invariant Lagrangian, the CP conjugate of triplet $\Sigma_{Rj}$ is defined as
 \begin{equation}
{\Sigma_{Rj}}^c=i\sigma_2\,C\,\overline{\Sigma_{Rj}}^{T}\,i\sigma_2 =
\left(\begin{array}{cc}
         {\Sigma^0_{Rj}}^c/\sqrt{2} & {\Sigma^-_{Rj}}^c \\
         {\Sigma^+_{Rj}}^c   & -{\Sigma^0_{Rj}}^c/\sqrt{2}
        \end{array}\right)~.
 \end{equation}
 Nevertheless, the CP conjugates of the component fields follow the usual definition
 i.e. $\Sigma^0_{Rj} = C\,\overline{\Sigma^0_{Rj}}^T$ and $\Sigma^{\pm}_{Rj} =
 C\,\overline{\Sigma^{\pm}_{Rj}}^T$. Now, using both $\Sigma_{Rj}$ and
 ${\Sigma_{Rj}}^c$, we can construct a triplet for each generation as
 \begin{eqnarray}
 \Sigma_j &=& \Sigma_{Rj} + {\Sigma_{Rj}}^c \nonumber \\
 && = \left(\begin{array}{cc}
         \Sigma^0_j/\sqrt{2} & \Sigma^+_j \\
         \Sigma^-_j   & -\Sigma^0_j/\sqrt{2}
        \end{array}\right)\,,
 \end{eqnarray}
 where, the neutral parts $\Sigma^0_j = \Sigma^0_{Rj} + {\Sigma^0_{Rj}}^c$
 are the Majorana fermions while the charged parts 
$\Sigma^{\pm}_j = \Sigma^{\pm}_{Rj} + {\Sigma^{\mp}_{Rj}}^c$ are the Dirac fermions. In terms of newly defined fields $\Sigma^0_j$ and $\Sigma^{\pm}_j$, the Lagrangian given in Eq.\,(\ref{lagN}) takes the following
form \cite{Biswas:2018ybc}
\begin{eqnarray}
 \mathcal{L}_{\Sigma} &=&  
 i\,\overline{\Sigma^{-}_j}\,\slashed{\partial} \Sigma^{-}_j +
 \frac{i}{2} \overline{\Sigma^{0}_j}\,\slashed{\partial} \Sigma^{0}_j - M_{\Sigma_j} \overline{\Sigma^{-}_j} \Sigma^{-}_j
 - \frac{M_{\Sigma_j}}{2} \overline{\Sigma^0_j} {\Sigma^0_j}
 -g\left(\overline{\Sigma^{-}_j}\slashed{W}^{-}\Sigma^0_j + h.c.\right)
 \nonumber \\ 
 &&+ g \cos\theta_W \overline{\Sigma^-_j} \slashed{Z} \Sigma^{-}_j
 + g \sin \theta_W \overline{\Sigma^-_j} \slashed{A} \Sigma^{-}_j 
 - \sqrt{2}\left\{ 
 y^{\alpha j}_{\Sigma} \left(\frac{1}{\sqrt{2}}
 \overline{\nu_{L\alpha}}P_R \Sigma^0_j +
 \overline{\ell_{L\alpha}}P_R\Sigma^{-}_j\right) {H^0}^{*}
 \right.
 \nonumber \\
 &&
 \left.
 -y^{\alpha j}_{\Sigma} \overline{\nu_{L\alpha}}P_R \Sigma^{+}_j H^{-}
 +\frac{y^{\alpha j}_{\Sigma}}{\sqrt{2}}\,
 \overline{\ell_{L\alpha}}P_R \Sigma^0_j H^{-}
 + h.c.\right\}
 \,.
\end{eqnarray}
Here, $H^-$ and $H^0$ are charged
and neutral components of the Higgs doublet $H$ respectively. As we mentioned above, all the elements in the first column of the
Yukawa coupling matrix are zero i.e. $y^{\alpha 1}_{\Sigma} =0$ for all
values of $\alpha$. This implies $\Sigma_{1}$ does not have Yukawa coupling with the SM lepton and Higgs. Thus, the neutral component of $\Sigma_1$, $\Sigma^0_1$
can be a suitable DM candidate. 

\section{Type III Seesaw explaining neutrino mass}\label{seesaw}
As discussed in Section \ref{model}, among the three triplet fermions, the 
$\mathbb{Z}_2$ odd one contributes to the DM phenomenology and
the other two heavier generations take part in neutrino mass generation
by the Type III seesaw mechanism. After the electroweak symmetry breaking, the neutrino mass matrix in the ($\nu_L$~~$\Sigma_{R}^{~c}$ ) basis 
takes the following form,
\begin{eqnarray}
\mathcal{L}_{NM} = - \frac{1}{2} \begin{pmatrix}
\overline{\nu_L} & \overline{\Sigma^c_{~L}}  
\end{pmatrix}
\begin{pmatrix}
0 & M_{D} \\
M^T_{D} &  M_{\Sigma}
\end{pmatrix}
\begin{pmatrix}
\nu^c_{~R} \\
\Sigma_{R}
\end{pmatrix}
+ {\it h.c.}
\label{neutrino-mass}
\end{eqnarray}
It is evident that the mass matrix, written in ($\nu_L$~~$\Sigma_{R}^{~c}$ ) basis, is a six dimensional square matrix if we consider three generations of light active neutrinos and three generations of fermion triplet. The constituent matrices $M_{D}, M_{\Sigma}$ are square matrices of order $3\times 3$.
In contrast to the usual type-I seesaw mechanism, the first column of $M_D$ is null since the Yukawa coupling of $\Sigma_1$ is forbidden by the $\mathbb{Z}_2$ symmetry. The three dimensional bare mass matrix of the triplet has been assumed to be diagonal with a hierarchical mass spectrum 
$M_{\Sigma_1} \ll M_{\Sigma_2} < M_{\Sigma_3}$.    
With the seesaw assumption i.e, $M_{\Sigma} \gg M_{D}$, 
we block diagonalize the mass matrix shown in Eq.\,(\ref{neutrino-mass}),
and get the following two eigenvalues (matrices) as,
\begin{eqnarray}
m_{\nu} & \simeq & - M_{D} M^{-1}_{\Sigma} M^T_{D} \nonumber \\
M_{\Sigma} & \simeq & {\rm diag}(M_{\Sigma_{1}}, M_{\Sigma_{2}}, M_{\Sigma_{3}})\,.
\label{diagonal-neutrino-mass}
\end{eqnarray}
In Eq.\,(\ref{diagonal-neutrino-mass}), $m_{\nu}$ is the active 
neutrino mass matrix and $M_\Sigma$ is the heavy fermion triplet mass matrix.
The effective neutrino mass matrix $m_\nu$ has to be diagonalized further in order to 
get the light neutrino mass eigenvalues and mixing angles.
In the present work, we are not concerned about the exact flavour
structure of the neutrino mass matrix and thus we are not going 
into the details of the diagonalization procedure. Existing studies
on Type-III seesaw leptogenesis have introduced a lower bound on the
next to lightest triplet mass from BAU bound, $M_{\Sigma_2} \gtrsim 3\times10^{10}$ GeV, which ensures sub-eV scale active neutrinos.
\section{Lightest triplet as DM candidate}\label{trip-dm}
The lightest triplet $\Sigma_1$ has been made stable by switching off its coupling to the SM particles through imposition of $\mathbb{Z}_{2}$ discrete symmetry. The three components $(\Sigma^+_1,\Sigma^0_1,\Sigma^-_1)$ of the triplet are degenerate at the tree level. However, the degeneracy is lifted when we consider radiative corrections \cite{Cirelli:2005uq}. In that case, the charged component becomes slightly heavier than the neutral one which makes the decay channel $\Sigma^{\pm}_1 \rightarrow \Sigma^0_1 + W^{\pm}(\rm virtual)$\footnote{These virtual $W^{\pm}$ decay into pions or leptons} allowed. Thus $\Sigma^0_1$ is the lightest stable particle in our particle spectrum, which has the potential to serve as a
viable dark matter candidate by reproducing the correct relic abundance
as predicted by the Planck collaboration \cite{Planck:2018vyg}. 
\subsection{Constraints on triplet DM}\label{cons-DM}
\begin{figure}[h!]
\centering
\includegraphics[angle=0,height=8.5cm,width=10.5cm]{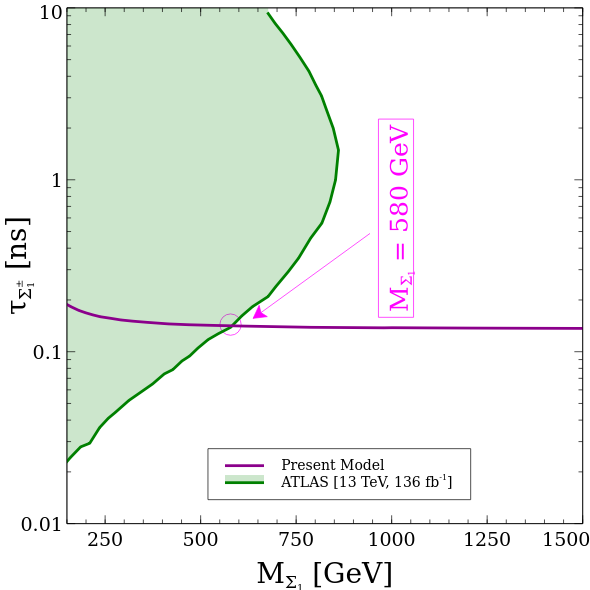}
\caption{Collider bound on the mass of the triplet fermion when the neutral
component is a DM candidate.} 
\label{DM-bound}
\end{figure}
The detailed bound on the triplet fermion DM has been addressed
in Ref.\,\cite{Belanger:2022gqc}. Considering the neutral component of
the triplet fermion as WIMP DM candidate, the bound can come from 
direct detection, indirect detection and collider searches. In the present model,
 the WIMP-nucleon scattering cross-section occurs at one loop level and gets suppressed
 accordingly \cite{Cirelli:2005uq}. As a result, the spin independent scattering
 cross-section ($\sigma_{\rm SI}$) lies at least one order of magnitude below
 the present bound from LUX-ZEPLIN (LZ) \cite{LZ:2022ufs} for $M_{\Sigma_1} = 1$ TeV \cite{Biswas:2018ybc}.
 Moreover, a further suppression on $\sigma_{\rm SI}$  can arise due to
 the negative contribution of two loop gluon mediated processes \cite{Hisano:2011cs}.
 On the other hand, in the present model, our DM candidate ($\Sigma^0_1$) can
 only annihilate into $W^{+}W^{-}$ final state at the tree level\footnote{Coannihilations
 with charged partners, $\Sigma^\pm_1$ are not considered here as these are not
 relevant for DM indirect detection.}. These $W^\pm$ bosons eventually
 produce $\gamma$-rays after undergoing cascade processes involving pions. 
Accordingly, we have an existing bound on the $\langle{\sigma v}\rangle_{W^+W^-}$ from
gamma-ray telescopes like Fermi-LAT which is sensitive to $\gamma$-rays
in the range $\sim 20$ MeV to 300 GeV \cite{Fermi-LAT:2009ihh}. The recent
bound on $\langle{\sigma v}\rangle_{W^+W^-}$ by Fermi-LAT through
observation of $\gamma$-rays coming from dwarf spheroidal galaxies (dSph) \cite{MAGIC:2016xys}
allows a narrow region of triplet DM mass between $1000\,\,{\rm GeV} \leq M_{\Sigma_1} \lesssim 1500$ GeV.
For $1500\,\,{\rm GeV}\lesssim M_{\Sigma_1} \lesssim 2700$ GeV, there is a
rapid growth in $\langle{\sigma v}\rangle_{W^+W^-}$ due to the Sommerfeld
enhancement \cite{Hisano:2003ec,Hisano:2004ds} and thus is ruled out from non-observation
of excess gamma flux over the known astrophysical background. However,
there is a more recent analysis in Ref.\,\,\cite{Leane:2018kjk}
where\footnote{In the said reference the authors have
argued that DM in the mass range $20 - 10^{5}$ GeV is fully safe from
the indirect detection searches (when DM particle under consideration
is produced by the thermal freeze-out mechanism.). Here, we mention this result just as a general observation. However, it has not been used in our analysis.}
the authors have derived a model independent bound on the total
annihilation cross-section of dark matter after taking into account
the bounds from Planck, Fermi-LAT and AMS. 

The bound on $M_{\Sigma_1}$ from collider searches has a similar
search strategy as the SUSY charginos. As the mass difference 
 between charged and neutral components, $\Delta M = M_{\Sigma^{\pm}_{1}}
 - M_{\Sigma_{1}} \simeq 166$ MeV \cite{Cirelli:2005uq}, is greater than
the mass of charged pion, $\Sigma^{\pm}_{1}$ decays predominantly
into $\Sigma^0_{1}$ and $\pi^{\pm}$  with approximately $98\%$
branching ratio. Due to the small mass gap $\Delta M$, the
produced pion leaves a very short track at the collider.
This kind of missing track searches put an upper bound
 on the triplet fermion mass and the current bound is $M_{\Sigma_{1}} > 580$ GeV from
 ATLAS data with a luminosity of 136 $fb^{-1}$ \cite{ATLAS:2022rme} (as shown
 in Fig.\,(\ref{DM-bound})). Moreover, there is a future projection of  
 HL-LHC from $3\,\,ab^{-1}$ luminosity data which
 will probe the triplet fermion mass up to $870$ GeV \cite{Dainese:2019rgk}.  
In this context it should be mentioned that 
considering the bound given
in Ref.\,\,\cite{Leane:2018kjk}, the lower bound on the triplet 
DM mass shifts to $M_{\Sigma_1} \gtrsim 350$ GeV and in that case
the bound from the LHC will be important. Most importantly, the collider bound is hard to evade with the introduction of beyond SM physics for the present setup, but we can easily evade the indirect detection
bound by introducing an additional dark annihilation channel of DM.
For example, we can introduce a dark abelian
gauge group under which $\Sigma_1$ is charged and then DM can annihilate into dark gauge bosons 
and lower the indirect detection bound. Of course, in this kind of setup, we also need to 
take into account the gauge anomaly which can be addressed by introducing additional triplet fermion.
The additional fermion mass can be kept beyond the inflation scale which will isolate its participation 
in our analysis. On the other hand, we can choose its mass at the lower scale which will have an impact on our DM analysis. Quantitatively, we anticipate that this kind of addition to our setup will change 
our DM phenomenology minimally but qualitatively there will be no change in our conclusion. This is just an example in the context 
of the present work to evade the indirect detection bound if 
the present model is in conflict with the future measurement of $\gamma-$ray
data. It is to be mentioned that we have not implemented this type of
extension in our present study and it is left for future exploration.
Moreover, our leptogenesis study will remain unaffected because the triplet fermions which take part in leptogenesis are neutral under the dark abelian gauge group. For the lower mass range of DM, we have considered the collider bound on the DM mass to be stricter than the indirect detection bound, since the indirect detection bound can be evaded following the techniques discussed above. On the other hand for the higher mass range of DM, the indirect detection bound is more effective (or in other words hard to evade) due to the occurrence of the Sommerfeld enhancement peaks.
\subsection{Evolution of comoving number density in non-standard cosmology} \label{NS_cosmo}
As discussed earlier the neutral component of the lightest
triplet fermion, $\Sigma^0_1$, can be a very good DM candidate
when we consider the corresponding Yukawa couplings are absent due to
the imposition of $\mathbb{Z}_2$ symmetry. 
Corresponding Boltzmann equation governing the
evolution of DM comoving number density
(a ratio of number density over entropy density {\rm s})
is given by,
\begin{eqnarray}
\frac{d Y}{d z} = - \frac{M_{\Sigma_1}}{z^{2}} \frac{1}{3 H(T)} 
\frac{d\,{s}}{d T} \langle \sigma v \rangle_{\rm eff} \left( Y^{2} - Y^{2}_{eq} \right)\,, 
\label{BE}
\end{eqnarray} 
where $z = \frac{M_{\Sigma_1}}{T}$ is a dimensionless quantity
and ${s}(T) = g_{s}(T) \frac{2 \pi^{2}}{45} T^{3}$ with
$g_{s}(T)$ being the effective degrees of freedom ({\it d.o.f})
related to the entropy density ${\rm s}$. The comoving
density can also be expressed as $Y=\sum_i Y_i$ where $i$ runs for all the species in the
dark sector i.e., $\Sigma^0_1$, $\Sigma^+_1$ and $\Sigma^{-}_1$.
The thermal averaged effective annihilation cross-section (denoted as $\langle \sigma v \rangle_{\rm eff}$)
includes contributions from annihilation as well
as co-annihilations involving dark sector species. The general expression of  $\langle \sigma v \rangle_{\rm eff}$ is given by \cite{Griest:1990kh} 
\begin{eqnarray}
\langle \sigma v \rangle_{\rm eff} &=& 
\sum_{i\,j}\langle \sigma v \rangle_{ij}
\dfrac{n^{eq}_{\Sigma^i_1}}{n^{eq}_{\Sigma_1}}
\dfrac{n^{eq}_{\Sigma^j_1}}{n^{eq}_{\Sigma_1}} \,, \nonumber \\
&=& \sum_{i\,j} \langle \sigma v \rangle_{ij}
\dfrac{g_i\,g_j}{g^2_{\rm eff}} \left(1+\Delta_i\right)^{3/2}
\left(1+\Delta_j\right)^{3/2}
\exp\left[-x\,(\Delta_i+\Delta_j)\right]\,,
\label{sigmaveff}
\end{eqnarray}
where, $n^{eq}_{\Sigma^i_1}$ is the equilibrium number density of
the species $i$ and thus the total equilibrium number density is given by
$n^{eq}_{\Sigma_1}=\sum_i n^{eq}_{\Sigma^i_1}$.  The last step of the above equation can be
easily computed using the Maxwell-Boltzmann distribution with $g_i$
being the internal degrees of freedom of species $i$ and 
$\Delta_i =( M_{\Sigma^i_1}-M_{\Sigma_1})/M_{\Sigma_1}$
is the fractional mass
difference between the species $i$ and the lightest one which
is nothing but our DM candidate $\Sigma^0_1$. Therefore,
$\Delta_{i} = 0$ and $\Delta{M}/M_{\Sigma_1}$ for
$i=\Sigma^0_1$ and $\Sigma^{\pm}_1$ respectively where
$\Delta{M}=M_{\Sigma^{\pm}_1} -M_{\Sigma_1} \simeq 166$ MeV as mentioned earlier.
The quantity $g_{\rm eff}$ in the denominator is given by
\begin{eqnarray}
g_{\rm eff} = \sum_i
g_i \left(1+\Delta_i\right)^{3/2} \exp(-x\Delta_i)\,. 
\end{eqnarray}
The thermal averaged cross-section when
the species $i$ and $j$ are in the initial state
is denoted by $\langle \sigma v \rangle_{ij}$,
where $i=j=\Sigma^0_1$ represents DM annihilation while
all other channels are co-annihilations. In the
model under consideration, DM can only annihilate into $W^{+}W^{-}$
final state. On the other hand, there are several co-annihilation
channels like $\Sigma^0_1\,\Sigma^{\pm}_1 \rightarrow W^{\pm}Z$,
$\Sigma^{+}_1\,\Sigma^{-}_1\rightarrow W^{+}W^{-}$,
$\Sigma^{+}_1\,\Sigma^{-}_1\rightarrow f\bar{f}$ ($f$
is the SM fermion),
$\Sigma^{\pm}_1\,\Sigma^{\pm}_1 \rightarrow W^{\pm}W^{\pm}$
and $\Sigma^{\pm}_1\,\Sigma^{\mp}_1 \rightarrow W^{\pm}W^{\mp}$.
These co-annihilation channels have a great impact on the
freeze-out dynamics of DM as the mass gap between DM ($\Sigma^0_1$) and the next-to-lightest species ($\Sigma^{\pm}_1$) is around $166$ MeV only, which is less than $0.02\%$ of $M_{\Sigma_1} =1$ TeV.
The expressions of individual annihilation and co-annihilation cross-sections along with $\langle{\sigma v}\rangle_{\rm eff}$ in $s$-wave approximation
are given in Appendix \ref{appendix1}.\\

In the Boltzmann equation, $H(T)$ is the Hubble parameter
which depends on the total energy content of the Universe.
However, due to the unavailability of experimental data before
the BBN, the energy content of the Universe in that epoch
is not well known to us. In this work, we will exploit this
freedom by modifying the Hubble parameter before the BBN \cite{DEramo:2017gpl}.  In particular, we consider
an extra species $\phi$ that increases the total energy density
($\rho$) of the Universe by contributing dominantly before the BBN and
$\phi$ red-shifts faster than the radiation. As $\rho>\rho_{rad}$,
the Hubble parameter is also large compared to the standard case
of the radiation domination and thus the scenario is referred
to as the fast expanding Universe in the literature \cite{DEramo:2017gpl}.   
%
%
The expansion rate of the Universe is quantified by the Hubble parameter which is related to the total energy density of the Universe as,
\begin{eqnarray}
H(T) = \sqrt{\frac{8 \pi\,G\,\rho(T)}{3}}\,,
\label{hubble}
\end{eqnarray}
where $G$ is the Gravitational constant.
The total energy density is now the sum of two components, i.e,
\begin{eqnarray}
\rho(T) = \rho_{\rm rad} (T) + \rho_{\phi} (T)\,.
\label{total_rho}
\end{eqnarray}  
The first term ($\rho_{\rm rad}$) is the radiation contribution
to the energy density, which dominates the total energy budget
in the standard cosmology before the era of matter-radiation
equality ($T\simeq 0.8$ eV) and has the following form,
\begin{eqnarray}
\rho_{\rm rad}(T) = \frac{\pi^2}{30}\, g_{*}(T)\, T^4\,, 
\label{rho_rad}
\end{eqnarray}
where, $g_{\star}(T)$ is the effective relativistic {\it d.o.f} related
to the radiation energy density at the concerned temperature.
On the other hand, the second term in Eq.\,\eqref{total_rho},
representing energy density of the species $\phi$,
red-shifts as \cite{DEramo:2017gpl},
\begin{eqnarray}
\rho_{\phi} \propto a^{-(4+n)}\,,
\end{eqnarray}
with $n> 0$. Since $\rho_{\phi}$ red-shifts faster
($\rho_{\phi} \propto T^{4+n}$ as $a\propto T^{-1}$)
than the radiation, we can have a Universe dominated
by the energy density of $\phi$ in the pre-BBN era and
reduces to the standard picture at $T\leq T_r$.
Imposing the entropy conservation in a unit co-moving volume,
the energy density $\rho_{\phi}$ takes the following form \cite{DEramo:2017gpl},
\begin{eqnarray}
\rho_{\phi}(T) = \rho_{\phi} (T_{r}) \left( \frac{g_{s} (T)}{g_{s} (T_r)} \right)^{\frac{4+n}{3}} \left(\frac{T}{T_r} \right)^{4+n}\,,
\label{rho-phi}
\end{eqnarray}  
 where, the reference temperature $T_r$ is defined as
 the temperature where contributions of the radiation and the species $\phi$ becomes equal and for $T<T_r$ we have the usual radiation
 dominance. Finally, using Eqs.\,(\ref{rho_rad} and
 \ref{rho-phi}), the total energy density at any temperature
 can be written as
  \begin{eqnarray}
  \rho(T) &=& \rho_{\rm rad}(T) + \rho_{\phi}(T) \nonumber \\
  &=& \rho_{\rm rad}(T) \left[ 1 + \dfrac{g_*(T_r)}{g_*(T)}
  \left(\frac{g_{s}(T)}{g_{s}
  (T_r)}\right)^{\frac{4+n}{3}} \left(\frac{T}{T_r} \right)^{n} \right]\,,
  \nonumber \\
  &=& \rho_{\rm rad}(T) \left[ 1 + \left(\frac{g_{*}(T)}{g_{*}
  (T_r)}\right)^{\frac{1+n}{3}} \left(\frac{T}{T_r} \right)^{n} \right]\,.
\label{matter-contribution}
\end{eqnarray}   
In the last step, we consider $g_{*}(T) = g_s(T)$, which is valid
before the era of neutrino decoupling that takes place around
$T\sim 1$ MeV.
Therefore, for $T >> T_{r}$, $\rho(T)>>\rho_{\rm rad}(T)$ and
thus the Hubble parameter gets enhanced according to
Eq.\,\eqref{hubble}, which has a
great impact on the evolution of particles in
the early Universe. In this work, we will study
the impact of such non-standard cosmology on 
our DM candidate and also on the matter anti-matter asymmetry
of the Universe in the context of a typical BSM model where
the SM particle spectrum is extended by three additional
SU(2)$_{\rm L}$ triplet fermions. As we mentioned
earlier, this model potentially explains the three important
issues of the SM, namely the nature of dark matter,
the origin of tiny neutrino mass and the physics behind the observed asymmetry between matter and anti-matter of the Universe.
\subsubsection{Determining the freeze-out temperature
$T_f$}\label{mod-tf}
As discussed in the preceding section, the presence of
an extra species in the early Universe modifies the Hubble
parameter in the following manner,
\begin{eqnarray}
    H(T) &=& \sqrt{\frac{8 \pi^3\,G g_{*}(T)}{90}}\, T^{2} \left[ 1 + \left( \frac{g_{*}(T )}{g_{*}(T_{r})} \right)^{\frac{1+n}{3}} \left( \frac{T}{T_{r}}\right)^{n} \right]^{1/2}\,, \\
    &=& H_{\rm rad}(T)\, \left[ 1 + \left( \frac{g_{*}(T )}{g_{*}(T_{r})} \right)^{\frac{1+n}{3}} \left( \frac{T}{T_{r}}\right)^{n} \right]^{1/2} \,.
\label{hubble-full}
\end{eqnarray}
In terms of $z\,\, (= \frac{M_{\Sigma_{1}}}{T})$,
the Hubble parameter is given by
\begin{eqnarray}
    H(z) = H_{\rm rad}(z) f(z,z_r,n)\,, \label{mod-hubble}
\end{eqnarray}
where, the function $f(z,z_r,n)$ 
(following definition of $z$,  $z_{r} = \frac{M_{\Sigma_{1}}}{T_{r}}$)
is defined as
\begin{eqnarray}
 f(z,z_r,n) = \left[ 1 + \left(\frac{g_{*}(M_{\Sigma_1}/z)}{g_{*}(M_{\Sigma_1}/z_r)} \right)^{\frac{1+n}{3}}
 \left( \frac{z_r}{z}\right)^{n} \right]^{1/2}\,,
\end{eqnarray}
and $H_{\rm rad}$ is the Hubble parameter for the standard radiation
dominated Universe. Since the reference temperature $T_r << T_f$, the
function $f(z,z_r,n)$ can be significantly larger than unity and this results in an enhanced Hubble parameter.
The increase in the Hubble parameter will make the freeze-out of
dark matter happen earlier in time. In the thermal WIMP paradigm,
DM gets thermally decoupled from the cosmic fluid once the
interaction rate of DM falls below the corresponding expansion rate of the
Universe and we can determine the freeze-out temperature by solving the following equation,
\begin{eqnarray}
    H(T) = n^{eq}_{\Sigma^0_1} \langle  \sigma v \rangle_{\rm eff} \,\,{\rm at}\,\, T = T_{f}\,,
\end{eqnarray}
where, $n^{eq}_{\Sigma^0_1}$ and $\langle{\sigma v}\rangle_{\rm eff}$ are already defined
earlier in Sec.\,\,\ref{NS_cosmo}.
Upon simplifying\footnote{This simplification can be found in appendix\,\ref{no-density} ( Eq.\,\eqref{zf-eqn}~-~Eq.\,\eqref{mod-zf} ).} the above equality using the expressions of Hubble parameter (Eq.\,\eqref{mod-hubble}) in non-standard cosmology, the equilibrium number density of the DM particle and
effective annihilation cross-section,  we get the following
transcendental equation for $z_f$,
\begin{eqnarray}
    z_{f} \simeq 28 - \ln\left[ \frac{M_{\Sigma_1}}{1000\,\,{\rm GeV}} \right] 
    - \ln\left[ f(z_f,z_r,n) \right] + \frac{1}{2} \ln\left[ z_{f} \right]\, ,
\label{iterative-eq}
\end{eqnarray} 
which can be solved iteratively to obtain the freeze-out
temperature $T_{f}\,\,(=\frac{M_{\Sigma_1}}{z_{f}}\,)$.
It is to be noted that Eq.\,\eqref{iterative-eq} contains an additional
term (with negative sign), compared to its standard counterpart, which depends on two non-standard parameters $n$, $T_r$. In our study, we have
focused on the regime where the non-standard contribution is large, which implies 
a significant negative contribution due to the $\ln\left[ f(z_f,z_r,n) \right]$ term and hence, $z_f$ in our case turns out to be much smaller than that of the standard radiation dominated case. The lower value of $z_f$ clearly signifies that our DM candidate freezes out at a higher temperature. So we can expect a higher relic abundance than the standard cosmological scenario.
It is worth mentioning here that in the standard scenario,
we could have lowered the $z_f$ value only 
by increasing the DM mass $M_{\Sigma_1}$ (i.e. by reducing
$\langle{\sigma v}\rangle_{\rm eff}$). However, the introduction
of non-standard cosmology, where we have two additional parameters $n,\,T_{r}$, 
can result in a similar kind of lowering in $z_f$. We will see later in this study that the reduction in $z_f$ due to the non-standard parameters will help us obtaining the correct value of relic density for the lower values of dark matter mass which is
not at all possible in the standard case of radiation domination.
\subsubsection{Analytical approximation for DM relic density}
We can solve the Boltzmann equation given in Eq.\,(\ref{BE})
analytically in the limit where effective degrees of freedom $g_{*}$ is
not changing significantly with temperature, which is valid as long as the
freeze-out occurs well above the era of QCD phase transition ($T\sim 100$ MeV).
The comoving number density after freeze-out ($z>z_f$ or $T < T_{f}$)
can be obtained easily as
\begin{eqnarray}
    Y_{\Sigma_{1}} \left( z  \right) \simeq
    \left[ \frac{1}{Y_{\Sigma_{1}}(z_f)} +
    \sqrt{\frac{\pi}{45}}M_{\Sigma_1}\,M_{pl}\, J\left(z,z_{r},n\right)\right]^{-1},
 \label{Comoving_DM}   
\end{eqnarray}
where $z = \frac{M_{\Sigma_{1}}}{T}$ as defined before, the function $J(z,z_{r},n)$ contains $\langle{\sigma v}\rangle_{\rm eff}$
and it also depends on the non-standard cosmology parameters in the following
way
\begin{eqnarray}
    J(z,z_{r},n) &=& \int^{z}_{z_{f}}
    \frac{g^{1/2}_{*}\,\beta (z^\prime)\,
    \langle {\sigma v} \rangle_{\rm eff}}{f(z^\prime,z_r,n)\,{z^\prime}^2}dz^\prime\,,
    \label{J-expression}\\
    && \hspace{-4.5cm} {\rm and} \nonumber \\
    \beta(z^\prime) &=& 1 - \frac{1}{3} \frac{d \ln  g_{*}(z^\prime) }{d\ln z^\prime} \,.\nonumber
\end{eqnarray}
Assuming $g_{*}$ does not change vigorously over the region of
integration (which also implies $\beta \simeq 1$) and the enhancement factor $f>>1$,
the above integration can be simplified as
\begin{eqnarray}
 J(z,z_r,n) \simeq \left(\frac{g_{*}(M_{\Sigma_1}/z_r)}
 {\bar{g_{*}}}
 \right)^{\frac{1+n}{6}}   
 \sqrt{\frac{\bar{g_*}}{z^{n}_r}}
 \int_{z_f}^z \dfrac{\langle {\sigma v} \rangle_{\rm eff}}{{z^\prime}^{2-n/2}} d z^\prime\,.
 \label{J-approx}
\end{eqnarray}
In the above expression for $J$, if we neglect
the first term within the brackets and set $z_r=1$
and $n=0$, we will recover the function $J$ for the
radiation dominated Universe. In the present model, we have
$s$ wave annihilations and the expression of
$\langle {\sigma v} \rangle_{\rm eff}$
under $s$ wave approximation is given in Appendix\,\ref{appendix1}. 
This allows us to compute the function $J$ analytically as
\begin{eqnarray}
\hspace{-0.7cm}
 J(z,z_r,n) &\simeq&
 \left(\frac{g_{*}(M_{\Sigma_1}/z_r)}
 {\bar{g_{*}}}
 \right)^{\frac{1+n}{6}} 
  \sqrt{\frac{\bar{g_*}}{z^{n}_r}} \,\langle {\sigma v} \rangle_{\rm eff}\,\times
  \left\{
\begin{array}{ccc}
\frac{2}{2- n}
 \left( \frac{1}{z^{1-n/2}_{f}} - \frac{1}{z^{1-n/2}}\right)
    & {\rm for } &  n \neq 2 \\
\ln(z / z_f) & {\rm for} &  n = 2
\end{array}
\right.
\end{eqnarray}
The presence of $z_r^{n/2}$ ($z_r>>1$) in the denominator makes
the function $J$ suppressed compared to the case of radiation domination
which eventually increases the comoving number density through
Eq.\,(\ref{Comoving_DM}) for a fixed value of $\langle {\sigma v} \rangle_{\rm eff}$.
The Eq.\,\eqref{J-approx} matches with Ref.\,\cite{DEramo:2017gpl} except the factor within the brackets
which the authors have considered $\sim \mathcal{O}(1)$.
Finally, once we have the co-moving number density then we can determine the DM relic density using the following expression,
\begin{eqnarray}
    \Omega_{\rm DM} h^{2} = 2.755 \times 10^{8} \left( \frac{M_{\Sigma_{1}}}{\rm GeV}\right)
    Y_{\Sigma_{1}} \left( z \gg z_f \right) \,.
\end{eqnarray}
\subsubsection{BBN Bound}\label{BBN-bound}
The extra species $\phi$ can contribute adequately to the total
energy density and hence to the Hubble parameter ($H$) during
the epoch of BBN depending on the parameters $T_r$ and $n$. The
extra energy\footnote{The extra energy should be red-shifted
at least equal to or faster than the radiation energy.
} that can be accommodated at the time of BBN over the
standard radiation background is estimated in terms of parameter
$N_{\rm eff}$ as
\begin{eqnarray}
\Delta{\rho} = \Delta{N_{\rm eff}}\,\rho_{\nu_L}\,,   
\end{eqnarray}
where, $\rho_{\nu} = \frac{7}{8}\, \alpha_{\nu/\gamma}\,
\rho_{\gamma}$ is the energy of a single species of
active neutrino at BBN and $\alpha_{\nu/\gamma} =
\left(\frac{4}{11}\right)^{1/3}$ is the ratio of active
neutrino to photon temperature after neutrino decoupling
around $T\lesssim$ 1 MeV\footnote{In our numerical analysis, we consider
$\alpha_{\nu/\gamma} =1$.}. The allowed value of $\Delta{N_{\rm eff}}$
based on standard cosmological history around the BBN is given
by \cite{Pitrou:2018cgg}
\begin{eqnarray}
\Delta{N_{\rm eff}} = N^{\rm BBN}_{\rm eff} - N_{\rm SM}
= 0.374\,\,\text{at\,\,95\%\,\,C.L.}\,,
\end{eqnarray}
where, theoretical prediction of $N_{\rm eff}$ based on
the SM is $N_{\rm eff} = 3.046$ \cite{Mangano:2005cc, deSalas:2016ztq} and
$N^{\rm BBN}_{\rm eff}
= 2.88 \pm 0.27$ at 68\% C.L. \cite{Pitrou:2018cgg}. Now, assuming
the extra contribution from $\phi$ remains within the allowed limit gives
the lower limit of the transition temperature $T_r$ for a particular
value of index $n$
\begin{eqnarray}
T_{r} \geq \left( \frac{g_{s}(T_{\rm BBN})}{g_{s}(T_{r})} \right)^{\frac{4+n}{3 n}} \left( \frac{g_{s}(T_{r})}{0.65\,\alpha^4_{\nu/\gamma}} \right)^{1/n}
\,\,{\rm MeV}\,.
\label{Tr-BBN-boune}
\end{eqnarray} 
The above bound has been used in all the scatter plots and
in particular a sharp effect of this bound can be observed in the $n-T_{r}$
plane of RP of Fig.\,\eqref{scatter-plot-3} and LP of Fig.\,\eqref{contour-paraspace}.
\subsection{Numerical results: viability of fermion triplet DM}\label{DM-num}
In this section, we are going to discuss the numerical findings of our study with special emphasis on evolution of DM relic density in non-standard cosmological background. We will proceed through graphical representation of numerical results followed by meticulous analysis of those figures in order to realize the impact of non-standard cosmology towards generation of adequate relic abundance.\\

Various line plots systematically analyze the possibility of earlier freeze-out of the DM in the modified cosmological scenario. The extent of non-standard contribution is quantified by the parameters $T_r,n$ which play a pivotal role in modifying evolution of the DM relic abundance and thus the frozen value of the relic density can be altered significantly by tuning the said non-standard parameters.
Larger relic abundance (compared to the standard case), in agreement with the current experimental data, can be obtained for a lower mass, of the dark matter, which is allowed by the indirect detection bound as well as the bound obtained from collider searches. This entire phenomenon has been depicted through numerous line plots in Sec.~\ref{plots-dm}.\\

Relic density depends on the model parameters ($M_{\Sigma_1},T_r,n$) in a bit complicated manner. To understand the nature of this dependence, when the parameters are varied simultaneously, we show scatter plots of frozen value of the relic density with mass of the triplet DM, where the non-standard parameters are shown by colour variation. Limiting values of these parameters ($M_{\Sigma_1},T_r,n$), required for successful prediction of DM relic density, have been estimated from similar scatter plots presented in the following Sec.~\ref{plots-dm}.
\subsubsection{Graphical analysis of numerical results} \label{plots-dm}
\begin{figure}[h!]
\centering
\includegraphics[angle=0,height=7.5cm,width=7.5cm]{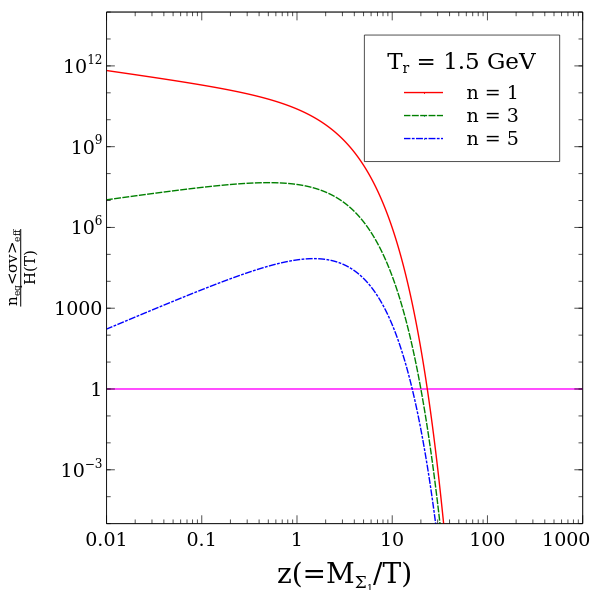}
\includegraphics[angle=0,height=7.5cm,width=7.5cm]{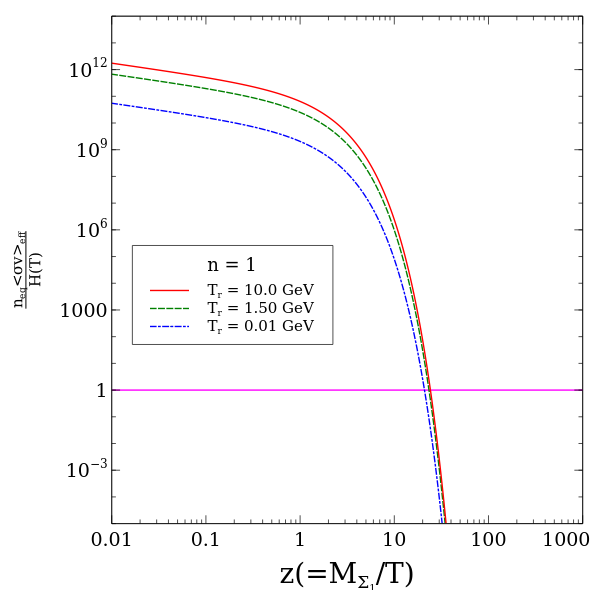}
\caption{LP and RP show the out of equilibrium check of DM for three
different values $n$ and $T_{r}$, respectively.} 
\label{line-plot-0}
\end{figure}

In the left panel (LP) of Fig.\,(\ref{line-plot-0}), we have shown variation in 
$z - \frac{n_{eq} \langle\sigma{}v\rangle_{\rm eff}}{H(T)}$ plane where 
$n_{eq} = \frac{g_{\Sigma} M^3_{\Sigma} K_{2}(z)}{2 \pi^2 z}$.
Here $K_{2}(z)$ is the modified Bessel function of second kind, $H(T)$ is Hubble parameter at temperature $T$ (defined  in Eq.\,(\ref{hubble})) and $\langle\sigma{}v\rangle_{\rm eff}$ is the effective annihilation
cross-section (defined in Eq.\,(\ref{effective-CS})). It is to be noted that the proposed DM candidate gets decoupled from the thermal plasma as soon as the corresponding reaction rate becomes less than the Hubble parameter at that instant. In the above figure, we have plotted the ratio of these two quantities, the out of equilibrium condition is satisfied whenever the value of this ratio falls below unity.
Thus it is clear from the figure (LP) that, as we increase the value of $n$,
the DM goes out of equilibrium for a lower value of $z$ (or equivalently for a higher value of temperature $T$). In other words, the DM freezes earlier for
higher values of $n$. Hence it can be inferred that with the increment of non-standard contribution, our DM goes out of equilibrium earlier and thus we will get higher DM abundance (as obtained from Eq.\,(\ref{iterative-eq})). 
On the other hand in the right panel (RP), we have shown
variation of the same quantity (as in LP) for three different values of $T_r$ while $n$ is kept fixed. From the figure (RP), it is clear that
lowering the value of $T_r$ causes earlier freeze-out, which implies a
greater relic abundance of the DM.
\begin{figure}[h!]
\centering
\includegraphics[angle=0,height=7.5cm,width=7.5cm]{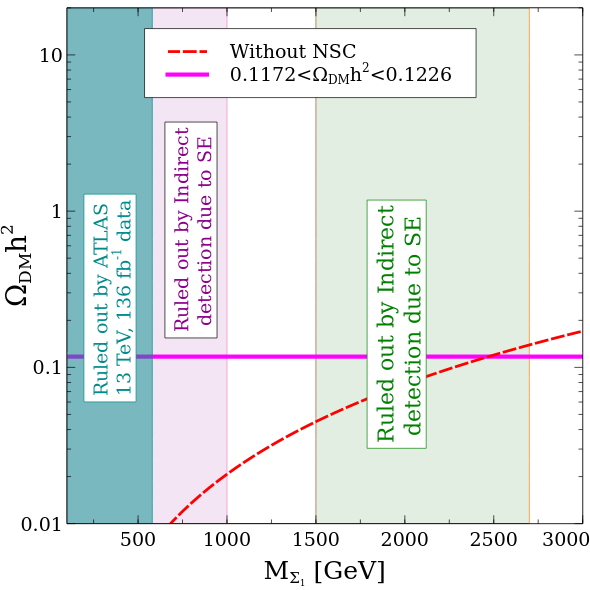}
\includegraphics[angle=0,height=7.5cm,width=7.5cm]{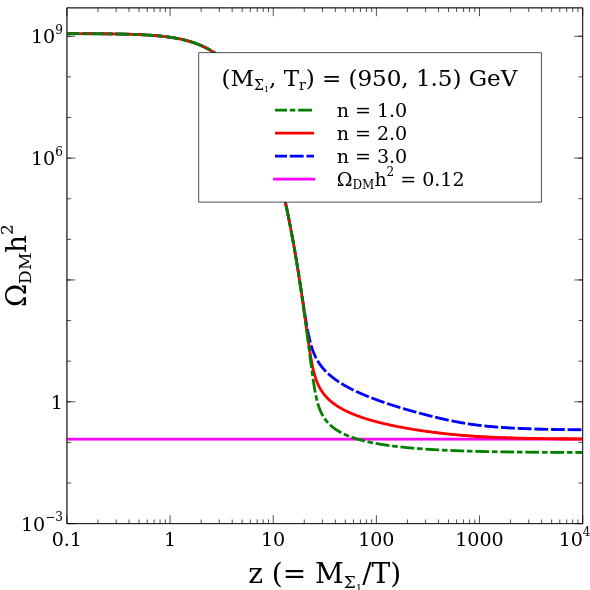}
\caption{LP: Variation of DM relic density with the DM mass for standard
cosmology (red dashed line) and non-standard cosmology (green solid line). RP: Variation of DM relic density with $z$ for three different values of $n$.} 
\label{line-plot-1}
\end{figure}

In the LP and RP of Fig.\,(\ref{line-plot-1}), we have shown line 
plots in the $M_{\Sigma_1} - \Omega_{\rm DM}h^{2}$ 
and $z-\Omega_{\rm DM}h^{2}$ planes, respectively. In the LP, we have
shown the dependence of relic density on DM mass where the red dashed 
line corresponds to the pure triplet model with standard cosmology \footnote{The line will be a little lower if we include the Sommerfeld factor in the relic density calculation and the correct value of DM relic density can be obtained for $2.7 \lesssim M_{\Sigma_{1}}\, ({\rm TeV}) \lesssim 3$  \cite{Hisano:2006nn}.} and
the green solid line represents the case with non-standard cosmological
background before the BBN. The green region is ruled out by Fermi-LAT due to the presence of the Sommerfeld enhancement spike around that specific mass range. The magenta region is also ruled excluded by Fermi-LAT and the cyan region is ruled
out by the missing track search at the LHC.
From the figure, it is clear that we have a very narrow window  
between $1000-1500$ GeV which is allowed after considering all the existing bounds mainly the indirect detection bound. As discussed in Section\,\,\ref{cons-DM}, the indirect detection bound can be evaded by 
introducing an additional annihilation mode of DM. Therefore from the lower mass end, 
the collider bound ($M_{\Sigma_{1}} > 580$ GeV) can be regarded as stricter than the indirect detection bound, whereas it is difficult to escape the indirect detection bound between 
$1.5$ TeV and $2.7$ TeV due to the presence of a spike (coming from the Sommerfeld enhancement factor) in the annihilation cross-section around the above mentioned mass range.
We can see that without non-standard cosmology it is very difficult to satisfy the 
DM relic density in the correct ballpark value. On the other hand, if 
we introduce non-standard cosmology before BBN then it helps to decouple the 
DM from the thermal bath earlier (can be seen from Eq.\,(\ref{iterative-eq})) 
and hence the higher value of relic abundance can be obtained. Therefore, we can clearly see that for $n=2$ and $T_{r} = 1.5$ GeV, we can easily satisfy DM relic density for mass
$M_{\Sigma_1} = 950$ GeV. On the other hand, in the RP, we have shown 
the DM relic density variation with $z$ for three different values of $n$. It is clear from the said figure that as we increase the value of $n$, the DM relic abundance is also enhanced. This is in agreement with the conclusion obtained from Fig.\,(\ref{line-plot-0}).
The most important information obtained from Fig.\,(\ref{line-plot-1})
is, in the non-standard scenario we can get the correct value of DM relic density 
for DM mass $M_{\Sigma_1} = 950$ GeV which is not achievable for the same DM mass within the framework of standard cosmology.
\begin{figure}[h!]
\centering
\includegraphics[angle=0,height=7.5cm,width=7.5cm]{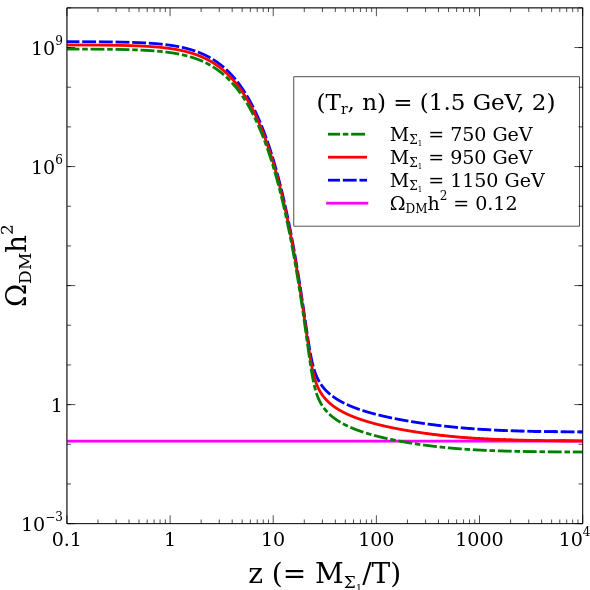}
\includegraphics[angle=0,height=7.5cm,width=7.5cm]{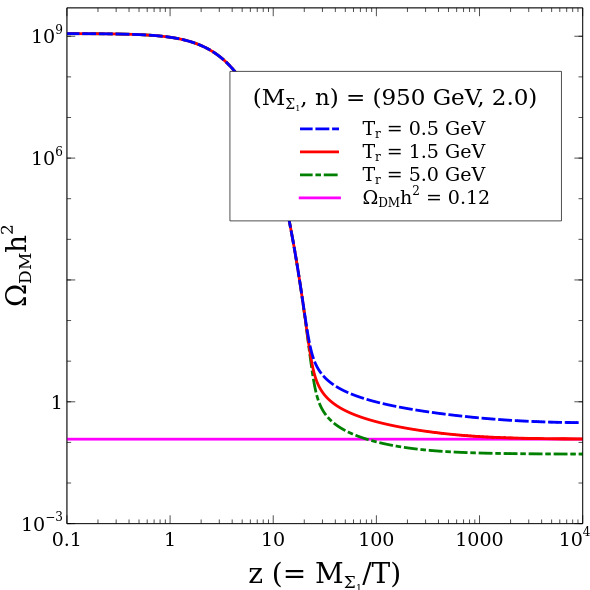}
\caption{LP and RP show the evolution of DM relic density with $z$
for three different values of DM mass $M_{DM}$ and reference temperature 
$T_{r}$, respectively.} 
\label{line-plot-2}
\end{figure}

In the LP and RP of Fig.\,(\ref{line-plot-2}), we have shown
the variation of DM relic density with $z$ for three different values of 
DM mass $M_{\Sigma_1}$ (while $T_r,n$ kept fixed) and $T_r$ (while $M_{\Sigma_1},n$ kept fixed ), respectively. It can be seen from the LP of the figure that
increment of DM mass enhances its relic density. 
This can be clearly understood from Eq.\,(\ref{th-cross-sec}) that 
thermal average cross-sections are inversely proportional to the square
of the DM mass and the relic density is inversely proportional to the thermal average 
cross-sections (Eq.\,\eqref{Comoving_DM}).
Therefore the DM relic density behaviour with mass is justified. The red solid line 
signifies that correct value of DM relic density can be obtained for  $M_{\Sigma_1}=950$ GeV while the non-standard 
parameters are kept at a value $T_{r} = 1.5$ GeV and $n = 2$. On the other hand, in the
RP, we have shown the DM relic density variation for three different values of $T_r$.
We have already discussed in the context of Fig.\,(\ref{line-plot-0}) that smaller values of 
$T_r$ implies more non-standard contribution in the DM production. The same behaviour of relic density 
is reflected in the RP of Fig.\,(\ref{line-plot-2}), i.e
if we lower the value of $T_r$, we will have more non-standard contribution in DM relic density and vice versa. The $T_r$ and $n$ dependence of DM production will be more vivid
when we discuss the scatter plots among the different parameters in the forthcoming figures. 
\\

In this part, we are going to show a few scatter plots among the model
parameters which affect the DM relic density significantly. The relevant
parameters are $M_{\Sigma_1}$, $n$ and $T_r$ which have been varied
in the following range (keeping in mind the upper bound on DM mass and BBN bound on $T_r$),
\begin{eqnarray}
500\,\,{\rm GeV} \leq M_{\Sigma_{1}} \leq 1500 \,\,{\rm GeV}\,,
10^{-4} \leq n \leq 5\,,\,\,{\rm and}\,\,10^{-3}\,\,{\rm GeV}
\leq T_{r} \leq 10^{3}\,\,{\rm GeV}\,.
\end{eqnarray} 
\begin{figure}[h!]
\centering
\includegraphics[angle=0,height=7.5cm,width=7.5cm]{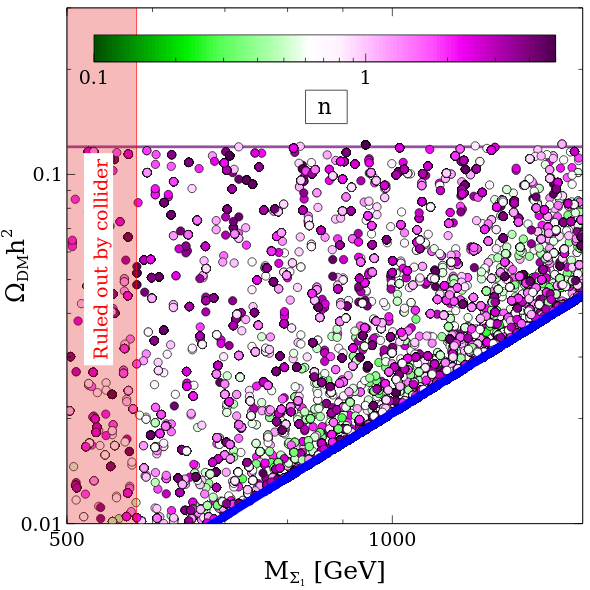}
\includegraphics[angle=0,height=7.5cm,width=7.5cm]{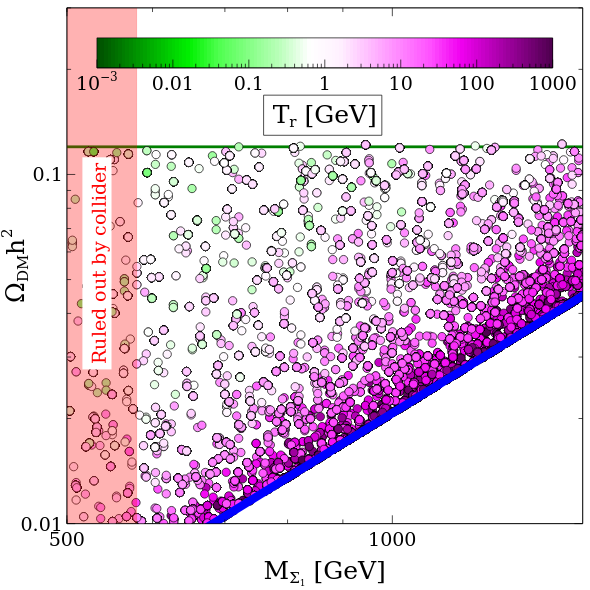}
\caption{LP: Scatter plot in the $M_{\Sigma_1} - \Omega_{\rm DM}h^{2}$ plane with 
the colour variation in the parameter $n$. In the RP, the same scatter plot 
but the color variation represent the parameter $T_{r}$. In both panels, the DM relic density has been assumed in the range $0.01$ to $0.12$.} 
\label{scatter-plot-1}
\end{figure}
In the LP of Fig.\,(\ref{scatter-plot-1}), we have shown scatter plot
in the $M_{\Sigma_1} - \Omega_{\rm DM}h^{2}$ plane allowing only those points for which
relic density lies within the range of $0.01-0.12$. 
Moreover, It should be noted that in the scatter plots we have chosen
strictly those combinations of $(T_r,n)$ which respect the BBN bound given in Eq.~(\ref{Tr-BBN-boune}). Basically, the plots of LP and RP (in Fig. \ref{scatter-plot-1}) both represent exactly the identical data-set composed of $(T_r,n,M_{\Sigma_1},\Omega_{\rm DM}h^{2})$. In LP variation of $n$ is shown by colour bar whereas in RP same technique is used to designate $T_r$.
The pink shaded\footnote{The pink shaded regions in the other plots imply the same thing.} region is already ruled out by the $136$ $fb^{-1}$ luminosity data of LHC. The blue solid line (in LP as well as RP of Fig.\,(\ref{scatter-plot-1})) corresponds to the DM relic density variation
with DM mass without any non-standard contribution. The $n$ value of any particular point in the plot can be retrieved from its colour. The horizontal colour bar shows continuous variation of $n$ with colour.
Before explaining this figure if 
we look at Eq.\,(\ref{matter-contribution}), we see that for $n > 1$ we can have a sizeable contribution, from the non-standard part, comparable to the standard part. On the other hand, if we take $T_r \gg 1$ 
the non-standard contribution, to the DM evolution, can be reduced to a large extent. In the figure, the lower sharp line (coinciding with the blue line) corresponds to the DM relic density for corresponding DM masses when non-standard contribution is negligible {\it i.e.} 
\[ \lim_{n > 1, T_{r}\gg 1}  \left(\frac{g_{*}(T)}{g_{*}(T_r)}\right)^{\frac{1+n}{3}} \left(\frac{T}{T_r} \right)^{n} \ll 1\,. \] 
It can also be seen from the figure that the green points correspond to 
$n < 1$ which gives more contribution than the standard 
contribution because of less suppression due to the aforementioned effect.
But as we have
seen in Fig.\,(\ref{line-plot-0}), the larger values of $n$ imply
more contribution from the non-standard cosmology and this is also consistent with 
the present figure because if we take a fixed value of $M_{\Sigma}$
and move along the ordinate then the DM relic density increases with the 
increase of $n$. In the RP of the same figure we have used similar kind of colour variation to bring out the nature $T_r$ dependence of relic density.
The dark-pink points, represent
higher values of $T_r$, give a smaller contribution in DM relic density
which is reflected in the region of negligible non-standard contribution. In the RP, for a fixed
value of $M_{\Sigma}$, if we move along the y coordinate we get
an increment in DM relic density which corresponds to lower values of
$T_r$ as represented by the colour variation. In both the figures, 
It has been checked explicitly that, for $n<0.4$ we can never achieve the correct value of DM relic density for the allowed DM mass range. We can't go below $n=0.4$ since it would result in a large value of $T_r$  (obtained from the bound of Eq.\,\eqref{Tr-BBN-boune}) which again tends to nullify the non-standard effect.
\begin{figure}[h!]
\centering
\includegraphics[angle=0,height=7.5cm,width=7.5cm]{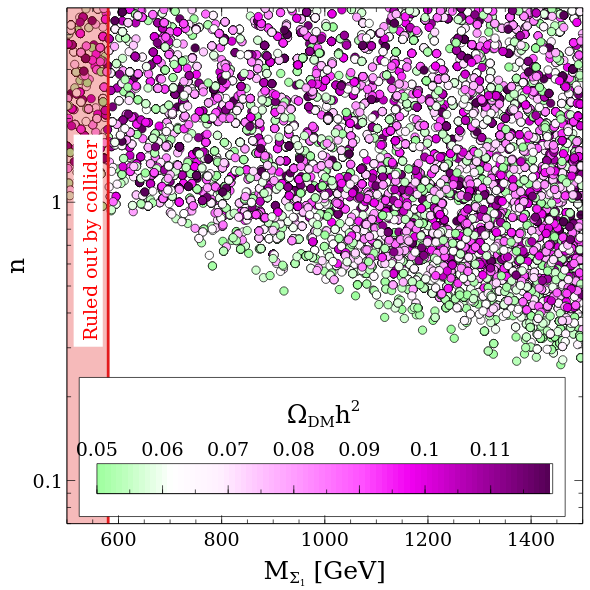}
\includegraphics[angle=0,height=7.5cm,width=7.5cm]{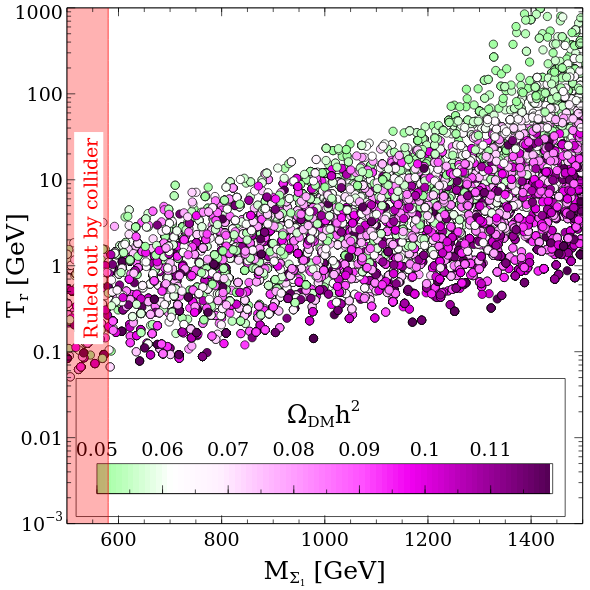}
\caption{LP and RP show the scatter plots in the 
$M_{\Sigma_1}-n$ and $M_{\Sigma_1}-T_{r}$ plane after demanding DM
relic density in the range $0.05$ to $0.12$. In both the panel the color variation shows the variation of DM relic density.} 
\label{scatter-plot-2}
\end{figure}

In the LP of Fig.\,(\ref{scatter-plot-2}), we have shown the scatter plot
in $M_{\Sigma_1} - n$ plane where the colour bar corresponds to the DM
relic density in the range $0.05-0.12$ and like
Fig.~(\ref{scatter-plot-1}), we have also demanded that all the points
satisfy BBN bound represented by Eq.\,\,(\ref{Tr-BBN-boune}).
We know from the LP of Fig.\,(\ref{line-plot-1}) that without non-standard cosmology $M_{\Sigma_{1}} = 1200$ GeV
corresponds to DM relic density $\Omega_{\rm DM}h^{2} \sim 0.03$. Thus from Fig.\,(\ref{scatter-plot-2}), it is clear that if the DM mass is in the range of $500 - 1200$ GeV and the non-standard parameter in the range
$0.1 \leq n \leq 0.4$, we can not have DM relic density around 0.05
for the whole range of $T_r$ considered in our work and therefore this
region is not allowed. The correct value of DM relic density can be achieved
for the whole range of DM mass if we are in the region $n > 1$ whereas the value of $n$
can be as low as $\sim 0.4$ if we take $M_{\Sigma_1} \sim 1500$ GeV. For the 
range $1200 <M_{\Sigma_1} ({\rm GeV}) < 1500$ and $0.1 < n < 0.4$, the non-standard is not
sufficient to produce the correct value of DM relic density. On the other
hand in the RP we have shown the variation of the parameters $M_{\Sigma} - T_{r}$
which produce DM relic density in the range of $0.05-0.12$.
It can be noticed from the figure that for the range $10 <T_{r}({\rm GeV})<1000$
and $500< M_{\Sigma_1} ({\rm GeV})< 1200$, the non-standard contribution is not sufficient
to raise the DM relic density up to $\Omega_{\rm DM}h^{2} = 0.05$. On the other
hand, if we take $T_{r}$ below $10$ GeV, we can always obtain the correct value of DM relic density. For the range $10  < T_{r}({\rm GeV})< 1000$ and 
$1200<M_{\Sigma_1}({\rm GeV})< 1500$, the non-standard contribution is not sufficient 
to have the correct value of DM relic density. 
\begin{figure}[h!]
\centering
\includegraphics[angle=0,height=7.5cm,width=7.5cm]{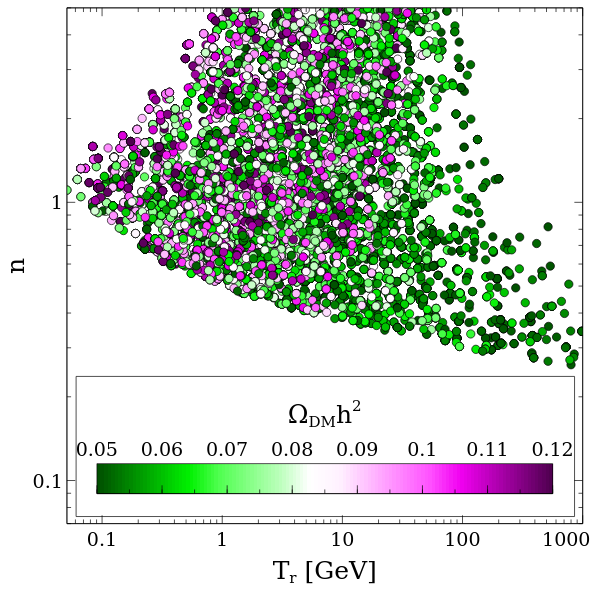}
\includegraphics[angle=0,height=7.5cm,width=7.5cm]{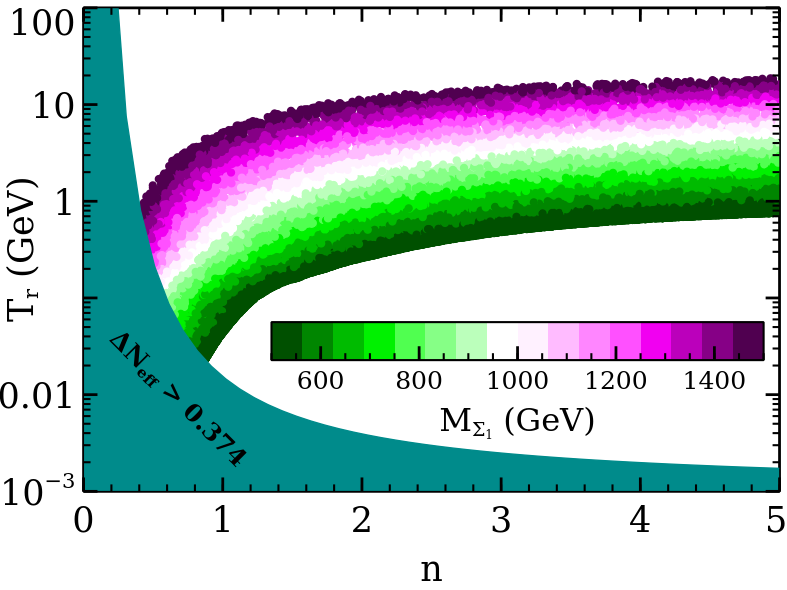}
\caption{Scatter plot in the $T_{r}-n$ plane after demanding DM relic density in the $0.05-0.12$. The colour variation represents 
DM relic density.} 
\label{scatter-plot-3}
\end{figure}

In the LP of Fig.\,(\ref{scatter-plot-3}), we have shown variation in the 
$T_{r} - n$ plane upon imposing the constraint that DM relic density (which is depicted in the colour bar) should lie within the range $0.05 - 0.12$. The top-left corner region represents an overabundant region, i.e. the
DM relic density $\Omega_{\rm DM}h^{2} > 0.12$. This can be easily understood since the region corresponds to a large non-standard contribution which drives the relic abundance towards a higher value. On the contrary,
we do not have any points in the top-right corner because that region has
less contribution from non-standard cosmology due to the high value of $T_r$. Thus our DM is severely underabundant in that region ($\Omega_{\rm DM}h^{2} < 0.05$). We also have discussed before that we can never get the correct value of relic density ($0.117\leq\Omega_{\rm DM}h^{2} \leq 0.123$) 
for $M_{\Sigma_1}\leq 1.5$ TeV if we are below $n = 0.4$. This figure also confirms the fact that the region below $T_r \lesssim 10$ GeV and $n > 0.4$
can reproduce the correct value of DM relic density. 
In the RP, we have shown the allowed region in $n-T_{r}$
plane that satisfies DM relic density in the $3\sigma$ range only. While doing
this scanning, we have varied all three variables $n$, $T_r$ and $M_{\Sigma_1}$ simultaneously. The corresponding variation of DM mass has been indicated
by the colour bar in the range $500\,\,{\rm GeV}\leq M_{\Sigma_1}\leq 1500$ GeV.
The region disallowed by BBN for producing excess $\Delta{N}_{\rm eff}$
(as given in Eq.\,\eqref{Tr-BBN-boune}) is shown by the dark cyan colour.
From this plot, one can easily notice that a larger DM mass is required
to satisfy $\Omega_{\rm DM} h^{2}$ in $3\sigma$ range when the effect
of non-standard cosmology is small i.e. for the lower values of $n$
and the higher values of $T_r$. Similarly, as we increase the DM mass,
the correct value of DM density can be achieved for a bit higher value
of $T_r$. However, once we go beyond $T_r \simeq 20$ GeV then one can never
achieve the correct DM relic abundance for any value of $n$ when
$M_{\Sigma_1} \leq 1.5$ TeV. In the later part of the article, we will
further constrain the $n-T_{r}$ parameter space once we impose
the BAU bound.
\section{Baryogenesis through Type-III seesaw leptogenesis}
\label{baryo-lepto}
The popular Type-I seesaw \cite{Minkowski:1977sc,GellMann:1980vs,Yanagida:1980xy,Mohapatra:1979ia} mechanism (SM + three generations of a gauge singlet heavy fermion), is primarily introduced to explain the non-zero mass of the active neutrinos, it has the potential to account for the observed baryon asymmetry of the Universe. Presence of Majorana type mass term in the Lagrangian ensures the lepton number violation by two units. The CP violating out of equilibrium decay of the singlet fermions (right handed neutrinos) to lepton Higgs pair gives rise to lepton asymmetry which gets converted into baryon asymmetry by sphaleron interactions \cite{Kuzmin:1985mm}. \\

If the said fermion singlet is replaced by SU(2)$_{\rm L}$ triplet fermion, the Majorana mass term and the Yukawa coupling term seems to be identical to that of the singlet except the fact that the Yukawa term now consists of three fold extra terms (which signifies the three components of the triplet) compared to its singlet counterpart. Unlike the singlets, the fermion triplets have gauge interactions. At
high temperature these gauge interactions are strong enough to promptly drive the triplet abundance to its equilibrium value. Thus in this case the final asymmetry is almost independent of the initial condition (i.e. whether we start from vanishing /equilibrium initial abundance).
\\

The preliminary objective of this work is to assess the viability of fermion triplet
DM (which is very tightly constrained by indirect detection searches of DM and collider searches)
in a non-standard cosmological background. The neutral component of the lightest triplet
acts as a DM candidate since it is odd under $\mathbb{Z}_2$ discrete symmetry, whereas the heavier generations of the triplet as well as the whole SM particle content are even under $\mathbb{Z}_2$.
The next to lightest triplet, having normal Yukawa coupling can decay to lepton, Higgs pair to produce the CP asymmetry required to explain the BAU. Existing studies \cite{Hambye:2003rt,AristizabalSierra:2010mv,Hambye:2012fh} in the literature dealing with hierarchical fermion triplet leptogenesis have already set a lower bound on the mass of the lightest decaying triplet to be $M_\Sigma \gtrsim  3\times 10^{10}$ GeV\footnote{ This is similar to the Davidson-Ibarra bound \cite{Davidson:2002qv} obtained in case of singlet, i.e $M_{N_1} \gtrsim 10^9$ GeV.}. 
It is to be noted that the said lower bound also depends on the effective neutrino mass 
parameter $(\Tilde{m})$ and its specific value mentioned above can be obtained for $\Tilde{m} \sim 0.001$ eV. It can be found from the concerned plot \cite{Hambye:2003rt,Hambye:2012fh} of allowed (by BAU bound) parameters in
$M_{\Sigma_2}-\Tilde{m}$ plane, that the lowest allowed value of $M_{\Sigma_2}$ lies near $10^{11}$ GeV over a wide range of $\Tilde{m}$. Thus in the remaining portion of our analysis $10^{11}$ GeV is referred to as the lower limit of $M_{\Sigma_2}$ for standard cosmology.\\

Now in context of non-standard cosmology, it will be an intriguing study to examine whether the above mentioned bound can be lowered further in current scenario. Implications of the baryon asymmetry bound on the free parameters $(T_R,n)$ of non-standard cosmology is also worth studying.

\subsection{CP asymmetry}\label{cpasy}
The fermion triplet $\Sigma_i = \Sigma_{R\,i} + {\Sigma_{R\,i}}^c$ is actually composed
of three components and in two by two notation it is expressed as
\begin{equation}
\Sigma_i =\left(\begin{array}{cc}
         \Sigma^0_i/\sqrt{2} & \Sigma^+_i \\
         \Sigma^-_i   & -\Sigma^0_i/\sqrt{2}
        \end{array}\right)~.
 \end{equation}
Among these three components $\left( \Sigma^+_i,\Sigma^0_i, \Sigma^-_i \right)$,
the neutral one, which is a Majorana fermion, mimics the behaviour of right handed
neutrino as in the case of Type-I seesaw. The charged states, $\Sigma^{\pm}_i =
\Sigma^{\pm}_{R\,i} + {\Sigma^{\mp}_{R\,i}}^c$, are the Dirac fermions. 
Now, due to the SU(2)$_{\rm L}$ symmetry, it can be argued that decay widths
(to lepton Higgs pair) of all three components should be equal and
are given by
\begin{equation}
\Gamma_{\Sigma^0_i}=\Gamma_{\Sigma^+_i}=\Gamma_{\Sigma_i^-}
=\frac{M_{\Sigma_i}}{8 \pi} |y^\dagger_{\Sigma} y_{\Sigma}|_{ii}, 
\end{equation}
where `$i$' is the generation index of the triplet. Thus, for our future calculations we will use the notation $\Gamma_{\Sigma_i}$ (omitting the `charge' index) for the decay width of a specific generation of the triplet irrespective of its charge. It is evident that the thermally averaged decay rate (Eq.\,(\ref{th-decay-rate})) becomes three times to that of the well known singlet right handed neutrino, since the equilibrium number density of the decaying triplet (considering the neutral and charged components) is simply three times of the right handed neutrino.
The CP asymmetry parameter \cite{Covi:1996wh}, quantified by the difference between decay rates of `heavy triplet to lepton Higgs pair' and its CP conjugate process, is expressed mathematically as
\begin{equation}
\epsilon_{\Sigma_i} = \frac{\Gamma\left( \Sigma_i \longrightarrow L~H\right )-\Gamma\left( \overline{\Sigma_i} \longrightarrow \bar{L}~\bar{H}\right )}{\Gamma\left( \Sigma_i \longrightarrow L~H \right )+\Gamma\left( \overline{\Sigma_i} \longrightarrow \bar{L}~\bar{H}\right )} ~,
\end{equation}
where the asymmetry parameter includes contribution from all three components of the triplet. The generic expression for CP asymmetry produced due to the decay of the next-to-lightest triplet ($\Sigma_2$) is given by \cite{Hambye:2003rt,AristizabalSierra:2010mv,Hambye:2012fh} 
\begin{equation}
\epsilon_{\Sigma_2} =\frac{3}{2} \left(\frac{M_{\Sigma_2}}{M_{\Sigma_3}} \right) 
\left( \frac{\Gamma_{\Sigma_3}}{M_{\Sigma_3}}\right)I_\Sigma \left( \frac{f_v - 2f_s}{3} \right) ~,
\label{cp_long}
\end{equation}
where
\begin{eqnarray}
&& I_\Sigma = \frac{\mathcal{I}m \left[ \left( y^\dagger_{\Sigma} y_{\Sigma} \right)_{23}^2 \right]}{|y^\dagger_{\Sigma} y_{\Sigma}|_{22} |y^\dagger_{\Sigma} y_{\Sigma}|_{33}}\nonumber\\
&& \hspace{-5.8cm} {\rm and} \nonumber \\
&& \frac{\Gamma_{\Sigma_3}}{M_{\Sigma_3}} = \frac{|y^\dagger_{\Sigma} y_{\Sigma}|_{33}}{8\pi} ~.
\end{eqnarray}
$f_v$ and $f_s$ are vertex and self energy correction functions given by
\begin{eqnarray}
&& f_v = \frac{{M_{\Sigma_3}}^2({M_{\Sigma_3}}^2 -{M_{\Sigma_2}}^2)}{({M_{\Sigma_3}}^2 -{M_{\Sigma_2}}^2)^2 + {M_{\Sigma_2}}^2 {\Gamma_{\Sigma_3}}^2 } ~, \\
&& f_s = 2 \frac{{M_{\Sigma_3}}^2}{{M_{\Sigma_2}}^2} \left[ \left( 1+ \frac{{M_{\Sigma_3}}^2}{{M_{\Sigma_2}}^2}\right)  \ln\left(1+\frac{{M_{\Sigma_3}}^2}{{M_{\Sigma_2}}^2} \right) -1 \right]~.
\end{eqnarray}
When the triplet mass spectrum is strongly hierarchical, effectively the asymmetry is produced (and survives till present epoch) solely due to the decay of lightest generation of $\mathbb{Z}_2$ even triplet, which is $\Sigma_2$ in this case. So we have to deal with $\epsilon_{\Sigma_2}$ only. For the sake of notational simplicity, from now on we will omit the subscript `$2$' in $\epsilon_{\Sigma_2}$. In strongly hierarchical limit both the above mentioned loop functions reduce to unity. Therefore the expression for CP asymmetry (Eq.\,(\ref{cp_long})) simplifies to
\begin{eqnarray}
\epsilon_\Sigma & = & \frac{3}{2}  \frac{M_{\Sigma_2}}{M_{\Sigma_3}} \frac{\Gamma_{\Sigma_3}}{M_{\Sigma_3}} I_{\Sigma}
\left(-\frac{1}{3} \right) \nonumber\\
                & = & - \frac{1}{16\pi}  \frac{M_{\Sigma_2}}{M_{\Sigma_3}} \frac{\mathcal{I}m \left[ \left( y^\dagger_{\Sigma} y_{\Sigma} \right)_{23}^2 \right]}{|y^\dagger_{\Sigma} y_{\Sigma}|_{22}}~,
\end{eqnarray}
which is exactly $1/3$ of the CP asymmetry obtained from the decay of hierarchical RH neutrinos in case of Type-I seesaw mechanism. Following the same arguments as that of the fermion singlet, the theoretically allowed maximum value of CP asymmetry turns out to be \cite{Hambye:2003rt,AristizabalSierra:2010mv,Hambye:2012fh} 
\begin{eqnarray}
\epsilon^{\rm max}_\Sigma &=& \frac{M_{\Sigma_2}}{8\pi v^2} (m_3 - m_1) \nonumber\\
                          &=& \frac{M_{\Sigma_2}}{8\pi v^2} \left( \frac{\Delta {m^2}_{31}}{m_3 +m_1} \right)~, \label{cp-asy-max}
\end{eqnarray}
where $m_i (i=1,2,3)$ are the light neutrino mass eigenvalues\footnote{Since two triplet fermions take part in the active neutrino mass generation so the mass 
of the lightest component among the active neutrinos would be zero, i.e., $m_{1} = 0$.} and $v(=246{\rm GeV})$ is the SM VEV. To get an estimate for the lowest value of $M_{\Sigma_2}$ that can produce adequate asymmetry to meet the observed value of baryon asymmetry, this above mentioned upper limit of CP asymmetry has to be used in the Boltzmann equations. A few earlier
studies \cite{Hambye:2003rt,AristizabalSierra:2010mv,Hambye:2012fh} dealing with fermion triplet leptogenesis have shown that the concerned lower limit on the mass of the lightest triplet lies near $10^{11}$ GeV.
\subsection{Boltzmann equations for leptogenesis in standard cosmology}
\label{boltzeqn}
The set of Boltzmann equations are used to track the evolution of lepton asymmetry with temperature as the Universe
gradually cools down. Let us define few important parameters, necessary for writing the Boltzmann equations. Instantaneous abundance of a particle species is conventionally expressed in terms of scaled number density
$Y_a(z)(=\frac{n_a(z)}{s(z)})$, where $n_a(z)$ is the number density of the particle species `$a$', $s(z)$ is the entropy density and $z$ is a dimensionless parameter ($z=M_{\Sigma_2}/T$)\footnote{We have used this definition of $z$ only for leptogenesis analysis. It should not be confused with the $z$ used in analysis of DM.} varying inversely with the temperature of the Universe. Since all three components of the triplet contribute equally to the asymmetry, it would be more convenient to handle the set of Boltzmann equations in terms of total number density of a specific triplet generation, i.e $n_{\Sigma_i}=n_{\Sigma^0_i}+n_{\Sigma^+_i}+n_{\Sigma^-_i}$. In the present case, the asymmetry is assumed to be generated solely due to the decay of $\Sigma_2$. 
Therefore we are concerned with the number density $n_{\Sigma_2}$ only\footnote{Expressions for equilibrium number densities of different particle species are given in Appendix\,\ref{no-density}.}. Since there is no other generation of the triplet involved in the Boltzmann equations, we can safely use the notation $n_{\Sigma}$ for $n_{\Sigma_2}$. The evolution of the triplet abundance and the corresponding $(B-L)$ asymmetry are governed by the following set of Boltzmann equations \cite{Hambye:2012fh}
\begin{eqnarray}
&& sHz \frac{d Y_\Sigma}{dz}= -\gamma_D \left( \frac{Y_\Sigma}{Y^{eq}_\Sigma} -1 \right) -2\gamma_A 
\left( \frac{{Y_\Sigma}^2}{{Y^{eq}_\Sigma}^2} -1 \right) ,\label{BE_sig1}\\
&& sHz \frac{d Y_{B-L}}{dz} = -\gamma_D \epsilon_\Sigma \left( \frac{Y_\Sigma}{Y^{eq}_\Sigma} -1 \right) -
\frac{Y_{B-L}}{Y^{eq}_l}\left( \frac{\gamma_D}{2} + \gamma^{sub}_\Sigma \right) \label{BE_BL1},
\end{eqnarray}
where $\gamma$ generically designates the reaction density\footnote{Detailed expression is given in Appendix\,\ref{reaction-density}.} of the process under consideration and the subscript `$D$' denotes the decay process, whereas `$A$' stands for $2$ $\rightarrow$ $2$ gauge boson mediated scattering processes. The $\gamma^{sub}_\Sigma$ term accounts for the contribution from $\Delta L =2$ scattering processes where the resonant contribution has already been subtracted. It is clear from the above set of Boltzmann equations that there is an extra term (compared to that of Type-I seesaw) involving $\gamma_A$ in the first equation. It appears since the fermion triplet can couple to the gauge bosons unlike the gauge singlet fermions.
The above set of Boltzmann equations can be rewritten in a convenient form as
\begin{eqnarray}
&& \frac{d Y_\Sigma}{dz}= -D(z) \left( Y_\Sigma(z)-Y^{eq}_\Sigma(z) \right)  -
{S_A}(z) \left( {Y_\Sigma(z)}^2-{Y^{eq}_\Sigma(z)}^2 \right) ,\label{BEsig2}\\
&& \frac{d Y_{B-L}}{dz} = -\epsilon_\Sigma D(z) \left( Y_\Sigma(z)-Y^{eq}_\Sigma(z) \right) -
\left(  W(z) +W^{sub}(z) \right)Y_{B-L}(z) \label{BEBL2},
\end{eqnarray}
where different decay and scattering terms are restructured and can be expressed in terms of some fundamental constants
and parameters of the model as follows. Decay $(D)$ term and gauge boson mediated scattering ($S_A$) terms are simplified as
\begin{eqnarray}
D(z)&=&\frac{\gamma_D(z)}{s(z)H(z)zY^{eq}_\Sigma(z)} \nonumber\\
    &=&\frac{n^{eq}_{\Sigma}(z)\Gamma_\Sigma \frac{K_1(z)}{K_2(z)}}{s(z)H(z)z\frac{n^{eq}_{\Sigma}(z)}{s(z)}}  \nonumber\\
    &=& \left(\frac{M_{pl}}{8\pi v^2 1.66 {g_\ast}^{1/2}} \right) \Tilde{m}z \frac{K_1(z)}{K_2(z)}
\end{eqnarray}
where $\Tilde{m}=8\pi\Gamma_\Sigma \frac{v^2}{{M_{\Sigma_2}}^2}$.
The thermal reaction rate of gauge boson mediated two body scattering process $\gamma_A$, involves integral of the reduced cross-section ($\hat{\sigma}$), of the relevant process, over the centre of mass energy of the initial particles, is computed as
\begin{eqnarray}
\gamma_A(z) &=& \frac{{M_{\Sigma_2}}^4}{64 \pi^4 z} \int \limits^{\infty}_{x_{min}} 
\sqrt{x} K_1(z\sqrt{x}) \hat{\sigma_A}(x) dx \nonumber\\
         &=& \frac{{M_{\Sigma_2}}^4}{64 \pi^4 z} I(z)\,,\\
 && \hspace{-5cm} {\rm where,} \nonumber \\
 I(z) &=& \int \limits^{\infty}_{x_{min}} 
\sqrt{x} K_1(z\sqrt{x}) \hat{\sigma_A}(x) dx \nonumber \,,
\end{eqnarray}
and the integration variable $x$ is defined as the ratio of centre of mass energy ($s$) to the square of  decaying particle's mass $(M_{\Sigma_2})$. The minimum value of centre of mass energy for a generic two body process $\left(a,b \rightarrow i,j,....\right)$ is given by $s_{min}={\rm max}\left[ (m_a +m_b)^2,(m_i+m_j+...)^2\right]$. The analytical expression for the reduced cross-section is given in Appendix\,\,\ref{cr-sec}. The gauge mediated scattering term ($S_A$) of
Eq.\,(\ref{BEsig2}) can be written as a function of the above mentioned integral as
\begin{eqnarray}
 S_A(z)&=&\frac{2\gamma_A(z)}{s(z)H(z)z{Y^{eq}_\Sigma(z)}^2}\nonumber\\
       &=&\left( \frac{\pi^2 g^{1/2}_\ast M_{pl}}{1.66\times 180 g^2_\Sigma} \right) \frac{1}{M_{\Sigma_2}}\left(\frac{I(z)}{z {K_2(z)}^2}  \right)\,,
\end{eqnarray}
where $g_\Sigma$ is the internal degrees of freedom of the Majorana type fermion triplet $\Sigma$.
Similarly the washout terms ( $W$ and $W^{sub}$ of Eq.\,(\ref{BEBL2}) ) due to inverse decay and $\Delta L =2$ scattering (off-shell) processes are given by
\begin{eqnarray}
 W(z) &=& \left( \frac{3 M_{pl}}{1.66\times 32 \pi v^2 g^{1/2}_\ast} \right) \Tilde{m} z^3 K_1(z)\,, \\
 W^{sub}(z) &=& \left( \frac{ M_{pl}}{1.66\times 64 \pi^2 g^{1/2}_\ast} \right) \frac{z^3}{{M_{\Sigma_2}}} \left \{  \int \limits^{\infty}_{x_{min}} 
\sqrt{x} K_1(z\sqrt{x}) \left(\hat{\sigma_s}(x) + \hat{\sigma_t}(x) \right) dx   \right\}  \nonumber\\
            &=&  \left( \frac{ M_{pl}}{1.66\times 64 \pi^2 g^{1/2}_\ast} \right) \frac{z^3}{{M_{\Sigma_2}}} \left( I_s(z) +I_t(z) \right)
\end{eqnarray}
with $I_s,I_t$ being integral of the $s$ channel\footnote{The resonance contribution has been properly subtracted for the $s$ channel process. Analytical expressions for $\hat{\sigma_s}(x),\hat{\sigma_t}(x)$ are given in Appendix\,\ref{cr-sec}. }
and $t$ channel (reduced) cross-sections.
\subsubsection{Baryon asymmetry and efficiency factor at present epoch}
At relatively lower (than the leptogenesis scale, i.e $T \ll M_{\Sigma_2}$) temperature (or equivalently $z \gg 1$ ) when the strength of lepton/baryon number violating interactions become negligibly small,
$Y_{B-L}$ freezes to a certain value and remains constant till the present epoch. Assuming vanishing  preexisting asymmetry (i.e $Y_{B-L}(z_i)=0$) the set of Boltzmann equations (Eqs.\,(\ref{BEsig2} and \ref{BEBL2})) can be solved up to a arbitrary value of $z$ to obtain
\begin{eqnarray}
 Y_{B-L}(z) &=& -\epsilon_\Sigma Y^{eq}_\Sigma(z \ll 1) \left \{ -\frac{1}{Y^{eq}_\Sigma(z \ll 1)}  
                \int \limits^{z}_{z_i} dz^\prime \frac{dY_\Sigma}{dz^\prime} e^{- \int \limits^{z}_{z^\prime} \left(  W(z^{\prime\prime}) +W^{sub}(z^{\prime\prime}) \right)dz^{\prime\prime} }\right\} \nonumber\\
             &=&  -\epsilon_\Sigma Y^{eq}_\Sigma(z \ll 1) \kappa(z)  ~, \label{kappa-int}
\end{eqnarray}
where the integral inside the curly bracket gives efficiency factor evaluated at any arbitrary $z$. If we assume $z_f$ to be the value of $z$ where $B-L$ asymmetry gets frozen, i.e
\begin{eqnarray}
  Y_{B-L}(z_f)           &=&    -\epsilon_\Sigma Y^{eq}_\Sigma(z \ll 1) \kappa(z_f) \nonumber\\
{\rm or},~  Y^f_{B-L} &=&    -\epsilon_\Sigma Y^{eq}_\Sigma(z \ll 1) \kappa^f ~.
\end{eqnarray}
The so called sphaleron process converts this frozen value of $B-L$ asymmetry to baryon asymmetry which is connected to $Y^f_{B-L}$ through the multiplicative sphaleron transition factor $a_{sph}=28/79$, i.e,
\begin{eqnarray}
Y_B(z) &=& a_{sph}Y_{B-L}(z) = -a_{sph}\epsilon_\Sigma Y^{eq}_\Sigma(z \ll 1) \kappa(z)~~~{\rm and}\\
Y^f_{B} &=&a_{sph}Y_{B-L}(z_f)=-a_{sph}\epsilon_\Sigma Y^{eq}_\Sigma(z \ll 1) \kappa^f~.
\end{eqnarray} 
The viability of our model can be tested through a comparison between the theoretically predicted value of baryon asymmetry ($Y^f_{B}$) with the observed
 one ($Y^{obs}_{B}$) which ranges between 
$8.9\times 10^{-11} <Y^{obs}_{B}< 9.2\times 10^{-11}$ at $95\%$ confidence level \cite{Planck:2018vyg}.
\subsubsection{Modification of the Boltzmann equations in non-standard cosmology}
It has already been mentioned in Sec.\,\ref{NS_cosmo} that, in non-standard cosmological scenario the Hubble parameter is modified due to the non-trivial contribution of the $\phi$ field in total energy density. It is well known that above the scale of electroweak (EW) symmetry breaking all the SM particles can be treated to be relativistic and thereby allows us to assume $g_\ast(T>T_{\rm EW})=g_{SM}\simeq106.75$. Since the processes relevant to generation and washout of lepton asymmetry take place much above the EW scale, $g_\ast(T)$ in that temperature regime behaves as a constant with a value $g_{SM}$. 
In present scenario the total energy density (Eq.\,(\ref{matter-contribution})) includes additional (apart from the well known radiation energy) temperature dependent terms involving non-standard parameters $T_r,n$.
It affects the Hubble parameter directly by introducing a steeper temperature dependence
as already shown in Eq.\,\eqref{mod-hubble}, i.e $H(z) = H_{\rm rad}(z) f(z,z_r,n)$ 
(with $f(z,z_r,n) >1$).
The particle physics inputs (i.e decay widths and scattering cross-sections) in the Boltzmann equations are totally unaffected by the changes incorporated in the cosmology.  
Therefore it can be stated unambiguously that in spite of introduction of modified cosmology, the basic structure of the Boltzmann equations
(Eqs.\,(\ref{BEsig2} and \ref{BEBL2})) remains unaltered, except the decay and scattering terms which are modulated by non-standard effects through a $z$ dependent term. Lets us first inspect the non-trivial modification in the decay term, i.e.
\begin{eqnarray}
 \Tilde{D}(z) &=& \frac{\gamma_D(z)}{s(z)H(z)zY^{eq}_\Sigma(z)}   \nonumber\\
      &=&\left( \frac{\gamma_D(z)}{s(z)H_{rad}(z)zY^{eq}_\Sigma(z)} \right)
      \frac{1}{f(z,z_r,n)} \,, \nonumber \\
      &=& \frac{D(z)}{f(z,z_r,n)}~. 
\label{D-NS}
\end{eqnarray}
Analogy between the structure of various decay and scattering terms guide us to realize that all the concerned terms ($D,S_A,W,W^{sub}$) will be modified (to $\Tilde{D},\Tilde{S}_A,\Tilde{W},\Tilde{W}^{sub}$ ) exactly in a similar manner through the common multiplicative factor $1/f(z,z_r,n)$, where
\begin{equation}
 f(z,z_r,n)=  \left[ 1+ 
 \left(\dfrac{g_{*}(M_{\Sigma_2}/{z})}
 {g_*(M_{\Sigma_2}/{z_r})}\right)^{\frac{1+n}{3}}
 \left( \frac{z_r}{z} \right)^n \right ]^{1/2} .
\end{equation}
Thus, in order to find the baryon (precisely $B-L$) asymmetry produced due to the decay of a heavy fermionic triplet in a non-standard cosmological scenario, we have to solve the same set of Boltzmann equations (Eqs.\,(\ref{BEsig2} and \ref{BEBL2})) with modified decay and scattering terms ($\Tilde{D},\Tilde{S}_A,\Tilde{W},\Tilde{W}^{sub}$).
\subsection{Analysis of numerical results: constraining the non-standard parameters}\label{numboltz}
The present study of fermion triplet leptogenesis in non-standard cosmology is mainly motivated to inspect the following intriguing issues: (i) possibility of producing observed asymmetry at some lower mass scale, (ii) whether baryon asymmetry bound can impose a more stringent constraint on the non-standard parameters which are already restricted by the DM relic density bound. It is worth mentioning that, since we are interested to find the lower limit (using BAU) on $M_{\Sigma_2}$, we should use theoretically possible maximum value of the CP asymmetry parameter. So we have used $\epsilon^{\rm max}_\Sigma$ (Eq.\,(\ref{cp-asy-max})) as the CP asymmetry throughout the analysis (for both standard and non-standard case).\\

Let us proceed with a brief discussion on the scheme of solution of Boltzmann equations (Eq.\,(\ref{BEsig2}), Eq.\,(\ref{BEBL2})) in non-standard cosmology (i.e, using $\Tilde{D},\Tilde{S}_A,\Tilde{W},\Tilde{W}^{sub}$). Although in standard cosmology scenario the decay/inverse decay and the scattering terms, have been expressed as explicit functions of $z$, they depend on other parameters like $\Tilde{m}$ and $M_{\Sigma_2}$ too. However in modified cosmological scenario (along with the standard ones) the non-standard parameters play a pivotal role in determining the decay/inverse decay and scattering terms. Therefore they ($\Tilde{D},\Tilde{S}_A,\Tilde{W},\Tilde{W}^{sub}$) become functions of $(z;\Tilde{m},M_{\Sigma_2},T_r,n)$ where $z$ is the integration variable and others are parameters of the model. It is worthwhile to mention that we have kept $\Tilde{m}$ fixed at a benchmark value $0.05$ eV throughout the analysis, whereas the other parameters are varied within suitable range to answer the questions raised in the first paragraph of this section. \\

\begin{figure}[h!]
\begin{center}
\includegraphics[width=7.2cm,height=7.2cm,angle=0]{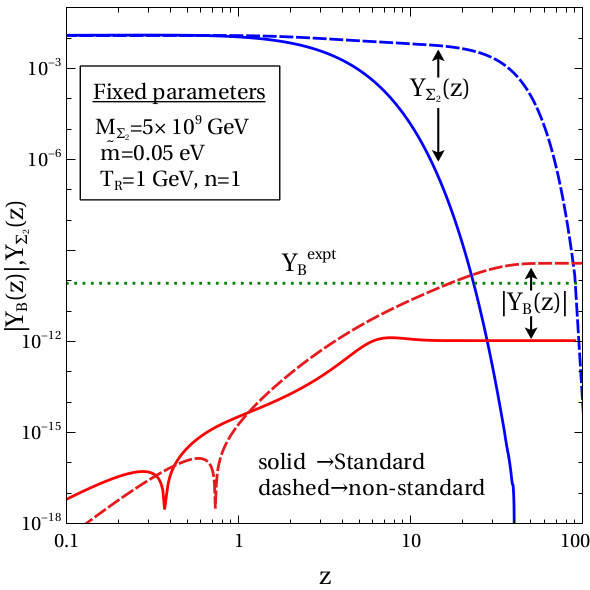}
\hspace{0.5cm}
\includegraphics[width=7.2cm,height=7.2cm,angle=0]{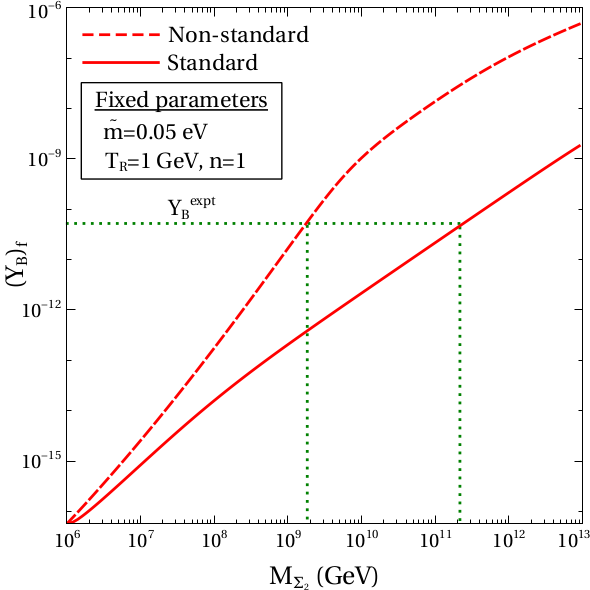}
\caption{{\bf Left:} Evolution of baryon asymmetry $Y_B$ (red line) and triplet abundance $Y_{\Sigma_2}$ (blue line) with $z$ assuming standard (solid line) as well as non-standard (dashed line) cosmology  for fixed benchmark value of set of parameters as $(M_{\Sigma_2}=5\times 10^9 {\rm GeV}, \Tilde{m}=0.05~{\rm eV}, T_r=1~{\rm GeV},n=1)$. Horizontal dotted line represents the experimental value of baryon asymmetry.  {\bf Right:} Variation of final (or frozen) value of baryon asymmetry ($Y^f_B$) with mass of the decaying triplet ($M_{\Sigma_2}$) for fixed values of the set of parameters $(\Tilde{m}=0.05~{\rm eV}, T_r=1~{\rm GeV},n=1)$. The particular points, where the vertical dotted lines have touched the abscissa, denote the lowest value of $M_{\Sigma_2}$ for which the BAU bound can be satisfied. }
\label{YB_z-and-YBf_M}
\end{center}
\end{figure}

At first we solve the Boltzmann equations for leptogenesis assuming standard radiation dominated cosmology for fixed value of triplet mass $M_{\Sigma_2}$. We show the variation of $Y_B$ with $z$ (in left panel of Fig.\,(\ref{YB_z-and-YBf_M})) to get an estimate of the specific value of $z$ above which the asymmetry freezes. The same exercise is repeated taking into account the presence of non-standard terms in the Boltzmann equations. The evolution of $Y_B$ is shown in the same plot together with its standard counter part. 
The figure (left panel, Fig.\,(\ref{YB_z-and-YBf_M})) vividly shows that, for chosen set of parameters the final asymmetry indeed freezes to a higher value in non-standard case. To investigate about the nature of dependence of final asymmetry ($Y^f_B$) on the decaying triplet mass ($M_{\Sigma_2}$), we solve the Boltzmann equations for a continuous range of $M_{\Sigma_2}$. The final asymmetry (for standard as well as non-standard) is plotted (right panel Fig.\,(\ref{YB_z-and-YBf_M})) against $M_{\Sigma_2}$, while the other parameters ($\Tilde{m},T_r,n$) are kept fixed at a bench mark value $(\Tilde{m}=0.05~{\rm eV}, T_r=1~{\rm GeV},n=1)$. As anticipated, $Y^f_B$ for
both the plots shows monotonically increasing behaviour. However, the non-standard curve (denoted by dashed line) shows a steeper dependence, thereby satisfying the baryon asymmetry bound at a much lower (approximately $2$ orders) value of $M_{\Sigma_2}$ compared to the standard one. So we have got the answer of our first question, i.e the lower bound on $M_{\Sigma_2}$ indeed decreases in presence of non-standard effect.\\

\begin{figure}[h!]
\begin{center}
\includegraphics[width=7.2cm,height=7.2cm,angle=0]{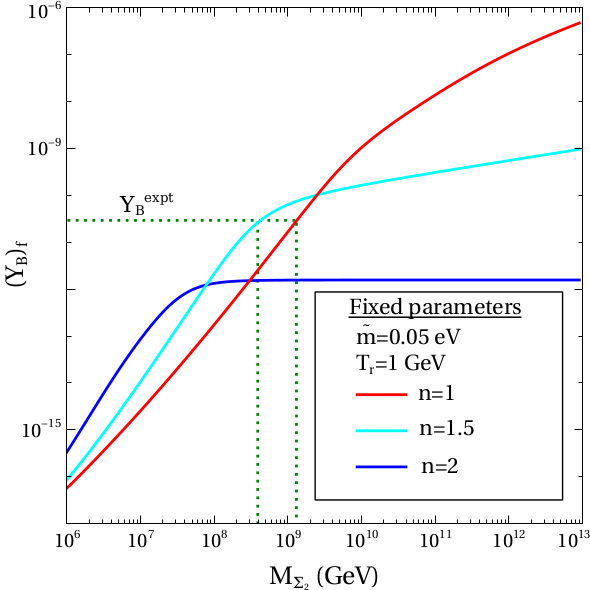}
\hspace{0.5cm}
\includegraphics[width=7.2cm,height=7.2cm,angle=0]{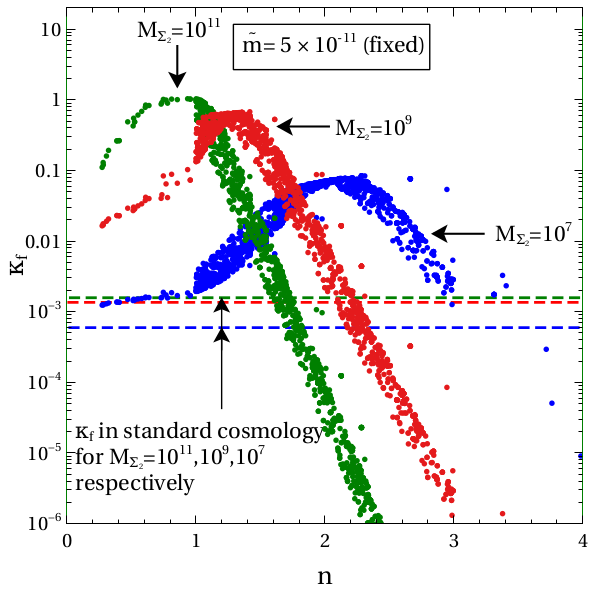}
\caption{{\bf Left:} Variation of final (or frozen) value of baryon asymmetry ($Y^f_B$) with mass of the decaying triplet ($M_{\Sigma_2}$) for three benchmark values of non-standard parameters $n$ while $\Tilde{m}$ and $T_r$ kept fixed throughout. The lower limit on $M_{\Sigma_2}$ decreases up to $n=1.5$.  {\bf Right:} Dependence of efficiency factor $(\kappa^f)$ on the non-standard parameter $n$ for three sample values of $M_{\Sigma_2}$. Horizontal dashed lines represent the corresponding efficiency factor in standard cosmology (obviously it does not depend on $n$). $M_{\Sigma_2}$ and $\Tilde{m}$ are in GeV unit.}
\label{YBf-kf}
\end{center}
\end{figure}

Let us recall the relation between final value of baryon asymmetry and efficiency factor, i.e
\begin{equation}
 Y^f_B(M_{\Sigma_2},\Tilde{m},T_r,n)  = -a_{sph} \epsilon_\Sigma (M_{\Sigma_2}) Y^{eq}_\Sigma(z \simeq 0)\kappa^f(M_{\Sigma_2},\Tilde{m},T_r,n) \label{YB_kappa2}
\end{equation}
where we have shown explicit dependence on model parameters. In principle this functional dependence indicates that final asymmetry expected to vary with the non-standard parameters $(T_r,n)$ even if 
$M_{\Sigma_2}$ is kept fixed. Therefore the conclusion drawn towards the end of last paragraph may not be true for all combinations of $(T_r,n)$. To examine its validity for other (than those used to draw the Fig.\,(\ref{YB_z-and-YBf_M})) values of `$n$' we redraw the same plot (i.e $Y^f_B$ vs $M_{\Sigma_2}$) for three sample values of $n$ and show them in the left panel of Fig.\,(\ref{YBf-kf}). 
It is observed that the said lower limit on $M_{\Sigma_2}$ decreases with $n$ only up to a certain value of $n$ (which is $n=1.5$ in this case), further increment in $n$ results in a drastic fall of final asymmetry which implies increment of the lower bound of $M_{\Sigma_2}$. This observation allows us to state  unambiguously  that the said lower bound can not be lowered indefinitely by increasing $n$. There exists a critical value of $n$ (for each fixed value of $M_{\Sigma_2}$) beyond which the final asymmetry declines monotonically. A deeper understanding about the nature of efficiency factor is required to shed some light on this issue. Eq.\,(\ref{YB_kappa2}) clearly designates that the non-standard parameters only affect the efficiency factor, whereas the CP asymmetry parameter being solely dictated by the underlying particle physics model remains unaffected. Variation of efficiency factor ($\kappa^f$) with 
$n$ for three different fixed values of $M_{\Sigma_2}$ is depicted in the right panel of Fig.\,(\ref{YBf-kf}). The efficiency factors obtained using standard cosmology (for same three benchmark values of $M_{\Sigma_2}$)
have been shown in the same figure with horizontal dashed lines. It is to be noted that the efficiency factor is indeed enhanced in non-standard case, but above a certain value of `$n$' it again starts to decrease and may become even lesser than its standard counterpart. This peculiar behaviour of $\kappa^f$ can be explained by studying the effect of non-standard parameters (specially $n$) on the terms responsible for the creation and washout of asymmetry. The integral representation (Eq.\,(\ref{kappa-int})) of efficiency factor $\kappa$ clearly indicates its dependence on `rate\footnote{This rate is again proportional to Decay and scattering terms (Eq.\,(\ref{BEsig2}))} of change of $Y_\Sigma$' and `the integral over washout terms'. It can be understood from Eq.\,(\ref{D-NS}), that the decay term $\Tilde{D}$ (and also $\Tilde{S}_A,\Tilde{W},\Tilde{W}^{sub}$) decreases with increase in $n$. When $n$ is increased from zero, initially the effect of decrease in washout (effectively increase in the integral) surpluses decrease in the Decay and scattering terms. Thus the resultant effect becomes positive which implies increment in the efficiency factor. However, beyond a critical value of $n$ the decrement in the $\Tilde{D}$ and $\Tilde{S}_A$ dominates overwhelmingly driving the efficiency factor towards a lower value.\\

The remarkable feature of this study is the interrelation, between DM phenomenology and leptogenesis,
which is brought about by the non-standard parameters. These parameters have taken a crucial role
in producing the correct relic abundance within the experimentally allowed range of $M_{\Sigma_1}$
as well as playing a vital role in bringing down the lower limit on $M_{\Sigma_2}$ (required for successful prediction of BAU).
In this context, it should be mentioned that during the computation of DM relic density, the non-standard parameters
are varied within the range $10^{-3} \leq T_r({\rm GeV}) \leq 10$ and $0 \leq n \leq 5$ while the mass of
DM $(M_{\Sigma_1})$ is taken between $500$ and $1500$ GeV. Moreover, we have also imposed the bound
on DM relic density in $3\sigma$ range as obtained by the Planck collaboration
i.e., $0.1172 \leq \Omega_{DM}h^{2} \leq 0.1226$. For leptogenesis, we have solved the Boltzmann
equations for the entire range of $(T_r-n)$ values previously allowed by relic density bound,
while $M_{\Sigma_2}$ has been varied independently over a wide range $(10^6~-~10^{13})$ GeV. 
\begin{figure}[h!]
\begin{center}
\includegraphics[width=7.3cm,height=7.2cm,angle=0]{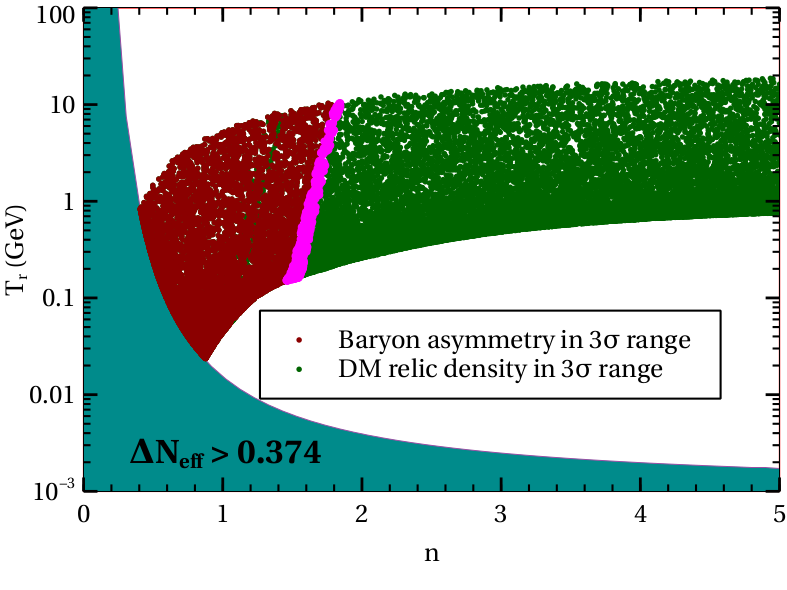}
\hspace{0.5cm}
\includegraphics[width=7.2cm,height=7.2cm,angle=0]{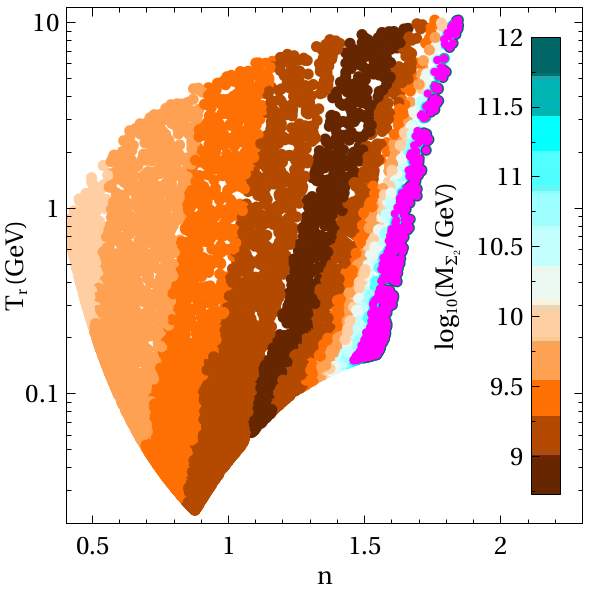}
\caption{ {\bf Left:} Non-standard parameter space $(T_r~-~n)$ allowed by relic density bound is denoted by deep green points whereas deep red points are those which satisfy both relic density as well as baryon asymmetry bound. The narrow magenta patch (which has actually covered few points in the dark red region) denotes those typical combinations of $(T_r,n,M_{\Sigma_2})$ which produce total energy density greater than the Planck limit. The area painted in deep cayan stands for the region excluded by BBN bound. The plot shows that the imposition of BAU bound reduces the allowed parameter space considerably. {\bf Right:} Allowed (by BAU bound) values of $M_{\Sigma_2}$ are represented by different colours in $(T_r~-~n)$ plane. The vertical colour bar stands for
specific numerical values of $M_{\Sigma_2}$ as denoted by different colours.
It can be seen that the magenta patch has overshadowed the allowed high mass region to some extent. This narrow region, although allowed by the baryon asymmetry bound, is in tension with the Planck constraint on the total energy density ($\rho_{\rm max} \sim 10^{66}$ GeV$^4$). However, the low mass region is completely safe from this bound.
The deep brown patch signifies the lowest achievable value of $M_{\Sigma_2}$, using non-standard cosmology which is $\lesssim 10^9$ GeV. }
\label{contour-paraspace}
\end{center}
\end{figure}
It can be inferred from the relevant plot (left panel, Fig.\,(\ref{contour-paraspace}))
that the parameter space (in $(T_r-n)$ plane) allowed by the relic density bound, shown by deep green points, has shrunk significantly to a smaller region consisting of deep red points upon imposition of second round of constraint, i.e. the BAU bound. For the sake of completeness we scan the whole allowed parameter space (by BBN bound + DM relic bound +  Baryon asymmetry bound) and compute the maximum possible value of total energy density ($\rho_{\rm tot}$) (using Eq.\,\,\eqref{matter-contribution}) in our model. The typical combinations of $(T_r,\,\,n,\,\,M_{\Sigma_2})$ which result in $\rho_{\rm tot}> \rho_{\rm max}$\footnote{The maximum allowed energy density\cite{Baumann:2009ds} of the Universe (as suggested by Planck data \cite{Planck:2018vyg}) after the inflation era is given by $\rho_{\rm max} \sim 10^{66}$ GeV$^4$.}, where $\rho_{\rm max}$ is the maximum allowed energy density of the Universe after inflation as dictated by the Planck data \cite{Planck:2018vyg}, are indicated by the magenta patch. Further in-depth studies related to leptogenesis,
like mass $(M_{\Sigma_2})$ distribution in $(T_r-n)$ plane, have to be done using this common parameter space designated by the
deep red patch in the above mentioned plot.\\

The meticulous analysis of numerical results, done so far conveys a clear message that dependence of final baryon asymmetry $Y^f_B$ on $M_{\Sigma_2}$ and other non-standard parameters is quite convoluted. Even the nature of dependence on a particular parameter (while others are kept fixed) is not always monotonic. Nevertheless, to get a comprehensive picture of the three dimensional parameter space (spanned by $M_{\Sigma_2},\,\,T_r,\,\,n$) allowed by recent experimental bound on BAU (as well as relic density bound), we depict\footnote{Only those combinations of $M_{\Sigma_2},\,\,T_r,\,\,n$ are considered which produces $Y^f_B$ within the experimental range.} ${\log_{10}}\left(M_{\Sigma_2}\right)$ with continuously varying colours
in ${\log_{10}}\left( T_r \right)-n$ plane
(right panel, Fig.\,(\ref{contour-paraspace})). The vertical colour bar 
contains all the required numerical information about the representative colours. The right most allowed region (deep cyan) corresponding to $M_{\Sigma_2} \sim 10^{12}$ GeV has been covered by the magenta patch which indicates that strict adherence with the Planck bound on total energy density will lead to exclusion of this narrow region from the allowed parameter space. However, the region associated with the lower values of $M_{\Sigma_2}$ remains absolutely unaffected (by the $\rho_{\rm max}$ bound).
The deep brown patch in the figure signifies that the lowest value of the decaying triplet mass, for which BAU bound can be satisfied, is less than $10^9$ GeV, while $1.0 \lesssim  n \lesssim 1.6$ and $T_r$ is less restricted, can take any value between $(1~-~10)$ GeV. Further increment/decrement in $n$ will shift lower bound on $M_{\Sigma_2}$ towards higher value.\\

\section{Summary and Conclusion}\label{summary}
\label{sandc}
In this work, we have considered the Type-III seesaw model
where the SM is extended by three SU$(2)_{\rm L}$ fermion triplets
to generate tiny Majorana masses for the SM neutrinos. However,
here we have imposed a $\mathbb{Z}_2$ symmetry on one of the triplet fermions $\Sigma_{R1}$ and accordingly, the neutral component of $\Sigma_{R1}$ becomes a stable dark matter candidate and one active neutrino remains massless. On the other hand, the heavier 
generations of the newly introduced triplet fermions can couple to the SM Higgs and 
leptons through Yukawa coupling, which in turn opens up the possibility to account for the 
observed baryon asymmetry of the Universe through CP violating out of equilibrium decay of
triplet fermions to lepton Higgs pair.\\

It is worthwhile to mention that although the stability of the proposed DM candidate ($\Sigma^0_{1}$) has been ensured by the imposed $\mathbb{Z}_2$ symmetry, it remains under-abundant
throughout the entire mass range (of $M_{\Sigma_1}$) allowed by the indirect and collider searches of DM.
Assuming the evolution of the relic abundance through standard radiation dominated cosmology, it can be shown that the correct relic density can be obtained for a fermion triplet DM of mass $\sim 2.3$ TeV which is ruled out unambiguously by the indirect searches of DM. To alleviate this problem or in other words to achieve the correct relic abundance for a relatively lower mass (allowed by experimental results) of the triplet, we resort to a typical form of non-standard cosmology where primordial (well above the BBN temperature) Universe expands faster than the standard picture.
In the present work we have shown that, indeed it is possible to produce the correct relic
abundance for a triplet mass as low as a few hundreds of GeV  while the values of the non-standard  parameters ($T_r-n$) have been chosen judiciously.  \\

From the point of view of particle physics, it seems that the dark sector and the Yukawa sector (responsible for asymmetry generation) are disconnected by the $\mathbb{Z}_2$ symmetry whereas the modified cosmology is bridging this gap through its parameters $T_r-n$.
These two parameters play a key role in getting correct relic abundance of the DM as well as
producing BAU in agreement with recent data. It has been found in our analysis that although the DM relic abundance constraint can be satisfied for high values of $n$ (up to the maximum value used for scanning the parameter space), the baryon asymmetry bound restricts it to be $\lesssim 1.8$.
On the other hand, the relic density constraint bounds $n$ from below with a limiting value of $0.4$. 
Thus it can be concluded that simultaneous satisfaction of DM relic density bound and BAU bound imposes a stringent restriction on the common parameter space spanned by the non-standard
parameters $T_r - n$. The exhustive analysis of leptogenesis leads us to another noteworthy result, i.e,
the lower bound on triplet mass (obtained from the observed BAU) can be brought down further by two orders of magnitude in comparison to the bound estimated using standard cosmology. 
\section*{Acknowledgements}
 The research of AB is supported by the National Research Foundation of Korea
 (NRF) funded by the Ministry of Education through the Center for Quantum Space
 Time (CQUeST) with grant number 2020R1A6A1A03047877 and by the Ministry of Science and
 ICT with grant number 2021R1F1A1057119. He also acknowledges the support
 from National Research Foundation grants funded by the Korean government
 (NRF-2021R1A4A2001897). M.C would like to acknowledge the financial support
 provided by SERB-DST, Govt. of India through the NPDF project PDF/2019/000622.  
 This work used the Scientific Compute Cluster at GWDG, the joint data center
 of Max Planck Society for the Advancement of Science (MPG) and University of
 G\"{o}ttingen. The authors would like to acknowledge the useful discussion with Laura Covi.

\appendix
\section{Thermal averaged cross-section} \label{appendix1}
In this Appendix, we have listed the expressions of all annihilation
and co-annihilation cross sections in the $s$-wave approximation as given
in Ref. \cite{Ma:2008cu},
\begin{eqnarray}
{\sigma v}_{\Sigma^0_1 \Sigma^0_1} \simeq \frac{2 \pi \alpha^2_L}{M^2_{\Sigma_1}}\,,
\,\,\,
{\sigma v}_{\Sigma^0_1 \Sigma_1^{\pm}} \simeq \frac{29 \pi \alpha^2_L}{8 M^2_{\Sigma_1}}\,,
\,\,\,
{\sigma v}_{\Sigma_1^{+} \Sigma_1^{-}} \simeq \frac{37 \pi \alpha^2_L}{8 M^2_{\Sigma_1}}\,,
\,\,\,
{\sigma v}_{\Sigma_1^{\pm} \Sigma_1^{\pm}} \simeq \frac{\pi \alpha^2_L}{M^2_{\Sigma_1}}\,. \nonumber\\ \label{th-cross-sec}
\end{eqnarray}    
The effective thermal averaged cross-section can be found easily using
Eq.\,\eqref{sigmaveff} as
\begin{eqnarray}\label{Appendix}
{\langle {\sigma v} \rangle}_{\rm eff} &=&
\frac{g^2_{\Sigma^0_1}}{g^2_{\rm eff}}
{\sigma v}_{\Sigma^0_1 \Sigma^0_1}
 +  2\,\frac{g^2_{\Sigma_1^{\pm}}}{g^2_{\rm eff}}
\left[{\sigma v}_{\Sigma^{\pm}_{1} \Sigma^{\pm}_{1}} +
{\sigma v}_{\Sigma^{+}_1 \Sigma^{-}_1} \right]
(1 + \epsilon)^{3} \exp(- 2 \epsilon x) \nonumber \\
&& + 4\,\frac{g_{\Sigma^0_1} g_{\Sigma^{\pm}_1}}{g^2_{\rm eff}}
{\sigma v}_{\Sigma^0_1 \Sigma^{\pm}_1} (1 + \epsilon)^{3/2} 
\exp(-\epsilon x)\,,
\nonumber \\
\label{effective-CS}
\end{eqnarray}
where $g_{\Sigma_1^0} = g_{\Sigma^{\pm}_1} = 2$,
$\epsilon = \frac{\Delta}{M_{\Sigma_1}}$,
$g_{\rm eff} = g_{\Sigma^0_1} + 2\, g_{\Sigma_1^{\pm}}
(1 + \epsilon)^{3/2}
\exp(-\epsilon x)$ 
and $\Delta = M_{\Sigma^{\pm}_{1}} - M_{\Sigma_1}
 \simeq 166$ MeV for $M_{\Sigma_1} \sim$ 1 TeV. So $\epsilon$ turns out to be negligibly small ($\sim 10^{-4}$). Thus in the limit $\epsilon \rightarrow 0$, the effective annihilation cross-section (Eq.\,\eqref{effective-CS}) can be approximated to be
 \begin{equation}
 {\langle {\sigma v} \rangle}_{\rm eff} \simeq \frac{111}{36} \frac{\pi {\alpha_L}^2}{{M_{\Sigma_1}}^2}~. \label{effective-CS-approx}   
 \end{equation}
\section{Equilibrium number density and freeze-out temperature}\label{no-density}
\paragraph{Number density of different species:}
The number density of fermion triplet (for each generation) in terms of temperature (or equivalently $z=M_\Sigma/T$) is given by
\begin{equation}
n^{eq}_{\Sigma}(T) = \frac{3}{4} \zeta(3) \frac{g_\Sigma T^3}{2\pi^2} \left(\frac{M_\Sigma}{T} \right)^2 K_2\left(\frac{M_\Sigma}{T} \right)~,
\label{nsig-eq}
\end{equation}
where $\zeta$ is the Riemann zeta function and $g_\Sigma$ is the internal degrees of freedom for the triplet fermion. In this case $g_\Sigma=2 + 4 = 6$, where the factor $2$ is for the Majorana fermion (neutral component) and the factor $4$ is for the Dirac fermion (charged component) respectively. Similarly number density of SM doublet leptons is 
\begin{equation}
n^{eq}_{l}(T) = \frac{3}{4} \zeta(3) \frac{g_l T^3}{\pi^2} ~,
\end{equation}
and that of photons is
\begin{equation}
n^{eq}_{\gamma}(T) =  \zeta(3) \frac{g_\gamma T^3}{\pi^2} ~,
\end{equation}
where $g_\gamma=g_l=2$ (since leptons can be treated to be massless above EW symmetry breaking scale).
\\

\noindent
Co-moving entropy density as a function of temperature is expressed as
\begin{equation}
 s(T) = \frac{2 \pi^2}{45} g_\ast T^3 ~,  
\end{equation}
where $g_\ast$ is number of effective relativistic degrees of freedom which turns out to be $\simeq 106.75$ if all Standard Model particles are assumed to be relativistic. 
\paragraph{Freeze-out temperature in non-standard cosmology:}
The dark matter gets thermally decoupled from the plasma at a specific temperature $T_f$ (or equivalently $z_f(=M_{\Sigma_1}/T_f)$ ) whenever the equality
\begin{equation}
  H(z_f) = n^{eq}_{\Sigma^0_1}(z_f) \langle  \sigma v \rangle_{\rm eff}   \label{zf-eqn}
\end{equation}
is reached. Using the modified Hubble (due to the presence of non-standard effects), i.e,
$H(z_f)=H_{\rm rad}(z_f) f(z_f,z_r,n)$, effective annihilation cross-section from Eq.\,(\eqref{effective-CS-approx}) and approximated form of the number density of the dark matter particle in the above equation we get
\begin{eqnarray}
\left(\frac{g_{\Sigma^0_1}}{\pi^2} {M_{\Sigma_1}}^3 {z_f}^{-3/2} e^{-z_f} \right) 
\left( \frac{111}{36} \frac{\pi {\alpha_L}^2}{{M_{\Sigma_1}}^2}  \right) =
\left(1.66 \sqrt{g_\ast(T_f)}\frac{1}{M_{pl}}\frac{{M_{\Sigma_1}}^2}{{z_f}^2}\right) f(z_f,z_r,n).
\end{eqnarray}
First we put numerical values of the constants $(g_{\Sigma^0_1},\alpha_L,M_{pl},g_\ast(T_f))$ in the above equation. Then a few steps of algebraic manipulations lead us to a much simplified form of the transcendental equation,
\begin{eqnarray}
{z_f}^{1/2} e^{-z_f} &\simeq& \left(6300\times10^{-19}\right) {M_{\Sigma_1}} f(z_f,z_r,n) \nonumber\\
{\rm or,~} z_f &\simeq& 35 + \frac{1}{2}\ln{z_f} -\ln{{M_{\Sigma_1}}} -\ln{f(z_f,z_r,n)}\nonumber\\ 
{\rm or,~} z_f &\simeq& 28 + \frac{1}{2}\ln{z_f} -\ln{\left({M_{\Sigma_1}}/1000\right)} -\ln{f(z_f,z_r,n)}, \label{mod-zf}
\end{eqnarray}
solving which we can get the freeze-out temperature $T_f=M_{\Sigma_1}/{z_f}$.
\section{Reaction densities of various decay, inverse decay, scattering processes}\label{reaction-density}
Let us consider a generic decay process $(\psi \rightarrow a+b+.....)$ and a scattering process $(\psi +\chi \rightarrow a+b+.....)$ where $\psi$ is the particle whose abundance is to be tracked. The space-time density of the generic scattering process in thermal equilibrium is expressed as
\begin{equation}
 \gamma (\psi +\chi \rightarrow a+b+.....)= \int d\Pi_\psi d\Pi_\chi f_\psi f_\chi \int d \Pi_a d\Pi_b.....\left( 2\pi \right)^4 \delta^4(p_\psi + p_\chi -p_a-p_b-........) |\mathcal{M}|^2 ,
\end{equation}
where $d\Pi_i =\frac{d^3p_i}{(2\pi)^32E_i}$ is the phase space factor, $|\mathcal{M}|^2$ is the squared transition amplitude and $f_i$ is the distribution function of the concerned particle species. For a decay process the above integral can be simplified to express the reaction density as
\begin{equation}
 \gamma (\psi \rightarrow a+b) =n^{eq}_\psi \frac{K_1(z)}{K_2(z)} \Gamma_\psi   ,
 \label{th-decay-rate}
\end{equation}
where $\Gamma_\psi$ is the decay width of $\psi$ to $a,b$ in the rest frame of $\psi$ and $n^{eq}_\psi$ is the equilibrium number density of the $\psi$ particle. For the scattering process. For the scattering process the thermally averaged reaction rate can be reduced to
\begin{equation}
 \gamma (\psi+\chi \rightarrow a+b) = \frac{T}{64 \pi^4}   \int \limits^{\infty}_{s_{min}}  ds \sqrt{s}K_1\left( \frac{\sqrt{s}}{T} \right) \hat{\sigma}(s), \label{scatt-int}
\end{equation}
where the integration is evaluated over the centre of mass energy $(s)$, whose minimum admissible value is either square of sum of the initial particle masses or final particle masses (which ever is greater), i.e $s_{min}=max\left[ (m_\psi +m_\chi)^2, (m_a +m_b)^2  \right]$.
Here $\hat{\sigma}(s)$ is the reduced cross-section which is related to the actual cross-section $(\sigma)$ of the concerned process through the function\footnote{$\lambda[x_1,x_2,x_3]=(x_1-x_2-x_3)^2-4x_2x_3$} $\lambda$ as
\begin{equation}
 \hat{\sigma} = 2s\lambda[1,\frac{{m_\psi}^2}{s},\frac{{m_\chi}^2}{s}]\sigma ~.   
\end{equation}
The integration variable $s$ in Eq.\,\eqref{scatt-int} can be replaced by a dimension-less one $(x=s/{m_\psi}^2)$ such that the reaction density of scattering can be expressed as
\begin{eqnarray}
\gamma (\psi+\chi \rightarrow a+b) = \frac{{M_\psi}^4}{64\pi^4 z} \int \limits^{\infty}_{x_{min}} \sqrt{x} K_1 (z\sqrt{x}) \hat{\sigma}(x)dx.
\end{eqnarray}
\section{Scattering cross-section of various processes}\label{cr-sec}
\paragraph{Gauge boson mediated scattering process:}
Reduced scattering cross-section for the gauge boson mediated processes\footnote{It includes $\Sigma,\Sigma \rightarrow $ all possible fermion doublets, Higgs, gauge bosons } is given by
\begin{equation}
 \hat{\sigma}_A =\frac{6 {g_{2L}}^4}{72\pi} \left[ \frac{45}{2} r(x) -\frac{27}{2}{r(x)}^3 -
 \left\{ 9\left({r(x)}^2-2 \right) + 18\left({r(x)}^2-1 \right)^2 \right\}\ln\left( \frac{1+r(x)}{1-r(x)}\right)   \right]   ,
\end{equation}
where $g_{2L}$ is the $SU(2)_L$ gauge coupling constant and $r(x)=\sqrt{1-4/x}$.
\paragraph{$\Delta L=2$ scattering processes:}
The off-shell part of the $\Delta L=2$ scattering cross-section (for $s$-channel process) is given by
\begin{align}
\hat{\sigma}_s (LH \rightarrow \Bar{L}H^\ast) &= \frac{\left( {y_{\Sigma_2}}^\dagger {y_{\Sigma_2}}  \right)^2_{22}}{4\pi}   \left[ 2 + x{D^2_s}^{sub} + \left( 2 -3x\frac{m_3}{\Tilde{m}_1}  \right)Re\left\{ D_s \right\} +3\frac{m_3}{\Tilde{m}_1}\left(x\frac{m_3}{\Tilde{m}_1}-2 \right) \right. \nonumber\\
 &-\left. \frac{2}{x}\ln(1+x)\left\{ 1+ \left( Re\left\{ D_s \right\} +3\frac{m_3}{\Tilde{m}_1} \right)(1+x) \right\}  \right]
\end{align}
where $D_s$ is the $s$ channel propagator given by
\begin{equation}
 D_s = \frac{1}{s-{M_{\Sigma_2}}^2 +i \Gamma_{\Sigma_2}M_{\Sigma_2}}   
\end{equation}
and ${D^2_s}^{sub}$ is the modulus square of `resonance contribution subtracted' $s$ channel propagator, expressed in terms of $D_s$ as
\begin{equation}
 {D^2_s}^{sub} = 1 -\frac{\pi}{\Gamma_{\Sigma_2}M_{\Sigma_2}|D_s|^2}\delta\left( s -{M_{\Sigma_2}}^2 \right)   ~.
\end{equation}
Cross-section for the $t$ channel process is given by
\begin{eqnarray}
 \hat{\sigma}_t (LL \rightarrow H^\ast H^\ast) = \frac{\left( {y_{\Sigma_2}}^\dagger {y_{\Sigma_2}}  \right)^2_{22}}{2\pi} \left[ \frac{3x}{2}\left(\frac{{m_3}^2}{{\Tilde{m}_1}^2} +\frac{2}{1+x}  \right)+  \left(3\frac{m_3}{\Tilde{m}_1} -\frac{3}{2+x} \right)\ln\left( 1+x \right) \right]~.
\end{eqnarray}
\newpage
\bibliographystyle{JHEP}
\bibliography{triplet-dm-lepto-27Feb.bib}
\end{document}